\documentclass[article]{JHEP3}
\usepackage{amsmath,epsfig}
\usepackage{amssymb,amsfonts}
\usepackage{feynmf}
\usepackage{epsfig}
\usepackage[vcentermath,enableskew]{youngtab}
\relax
\renewcommand{\theequation}{\arabic{section}.\arabic{equation}}
\def\be{\begin{equation}}
\def\ee{\end{equation}}
\def\bea{\begin{eqnarray}}
\def\eea{\end{eqnarray}}
\def\be{\begin{equation}}
\def\ee{\end{equation}}
\def\ie{{\it i.e.}~}

\def\pa{\partial}

\newcommand\fverb{\setbox\pippobox=\hbox\bgroup\verb}
\newcommand\fverbdo{\egroup\medskip\noindent%
                        \fbox{\unhbox\pippobox}\ }
\newcommand\fverbit{\egroup\item[\fbox{\unhbox\pippobox}]}
\newbox\pippobox
\def\ve{\varepsilon}

\def\e{\epsilon}
\def\h{\eta}
\def\ka{\kappa}
\def\z{\zeta}
\def\m{\mu}
\def\n{\nu}
\def\r{\rho}
\def\s{\sigma}

\def\t{\tau}
\def\y{\psi}
\def\sp{\;\;\;,\;\;\;}

\def\p{\partial}

\def\f{\varphi}
\def\a{\alpha}
\def\b{\beta}
\def\d{\delta}
\def\l{\lambda}
\def\g{\gamma}
\def\G{\Gamma}

\def\la{\langle}
\def\ra{\rangle}

\def\ba{\begin{eqnarray}}
\def\ea{\end{eqnarray}}

\def\bn{N_{\bullet}}
\def\bo{N_{\rm o}}

\def\hre#1#2{\href{http://arxiv.org/abs/#1/#2}{[ArXiv:#1/#2]}}

\def\nn{\nonumber}
\def\sla{\raise.15ex\hbox{$/$}\kern-.57em}
\title{Anomalies, Anomalous U(1)'s and generalized Chern-Simons terms}

\author{\large P. Anastasopoulos$^1$,  M. Bianchi$\ ^2$, E. Dudas$\ ^{3,4,5}$, E.
Kiritsis$\ ^{3,6}$\\
~\\
$^1$Department of Physics,
University of Athens,\\
Panepistimiopolis, Zografou- 157 84 Athens, GREECE\\
~\\
$^2~$Dipartimento di Fisica, \ Universit{\`a} di Roma \ ``Tor
Vergata''
$~$\\
I.N.F.N.\ -\ Sezione di Roma \ ``Tor Vergata''
$~$\\
Via della Ricerca  Scientifica, 1 - 00133 \ Roma, \ ITALY
\\
$~$\\
$^3~$CPHT, UMR du CNRS 7644, Ecole Polytechnique,\\
91128 Palaiseau, FRANCE\\
$~$\\
$^4~$LPT, UMR du CNRS 8627, Bat. 210, Univ. Paris-Sud,\\
91405 Orsay Cedex, FRANCE\\
$~$\\
$^5~$CERN Theory Division, CH-1211,
Geneva 23, SWITZERLAND\\
$~$\\
$^6~$Department of Physics, University of Crete, 71003 Heraklion,
GREECE}

\preprint{\hepth{0605225}\\
CPHT-RR017.0306, ROM2F/2006/10
LPT-ORSAY-06-34, CERN-PH-TH/2006-084}      

\abstract{A detailed analysis of anomalous U(1)'s and their
effective couplings is performed both in field theory and string
theory. It is motivated by the possible relevance of such
couplings in particle physics, as well as a potential signal
distinguishing string theory from other UV options. The most
general anomaly related effective action is analyzed and
parameterized. It contains St\"uckelberg, axionic and
Chern-Simons-like couplings. It is shown that such couplings are
generically non-trivial in orientifold string vacua and are $not$
in general fixed by anomalies. A similar analysis in quantum field
theories provides similar couplings. The trilinear gauge boson
couplings are also calculated and their phenomenological relevance
is advocated. We do not find qualitative differences between
string and field theory in this sector.}

\begin{document}


\section{Introduction and conclusions}
It took some time for gauge and gravitational anomaly cancellation
to take its place as a cornerstone in the building of theories of
the fundamental interactions, \cite{bim}. Anomaly cancellation
provides powerful constraints on chiral particle spectra. Gauge
anomalies are intimately related to the UV structure of quantum
field theories (QFTs). Their presence imply UV divergences that
cannot be renormalized. There are several types of anomalies that
plague gauge and gravitational theories. All of them are fatal in
the UV of a QFT. However, their structure can be different in
different cases. We therefore have (in 4d) pure non-abelian cubic
anomalies, mixed anomalies between non-abelian and abelian gauge
groups, as well as cubic abelian anomalies. In addition to this we
have mixed abelian-gravitational anomalies associated with the
trace of U(1) charges.

In string theory, the situation is slightly different. Closed string
theory has a UV regime protected by the stringy cutoff introduced by
the geometry of the closed Riemann surfaces. Modular invariance is
crucial in this. It is the same invariance that guarantees the
absence of irreducible (non-factorizable) anomalies. Reducible
anomalies can be cancelled via the the Green-Schwarz mechanism
\cite{gs}, and its generalizations. Mixed abelian-non-abelian
anomalies and cubic abelian anomalies are in this class.
Generically, the chiral fermionic spectra in string models are not
anomaly-free by themselves, but the gauge variation of their
contribution to the one-loop effective action is precisely cancelled
by antisymmetric tensor fields of various ranks which undergo
non-linear gauge transformations \cite{gsw}. In earlier perturbative
heterotic constructions, the Green-Schwarz mechanism involves only
one, universal, axion \cite{dsw}. There have been recent discussions of
heterotic compactifications in the supergravity limit, where the possibility
of several anomalous U(1) factors was pointed out, \cite{het}.
This matches the situation in orientifold vacua
\cite{orientifolds1,orientifolds2}, which contain
several axions or antisymmetric tensors
\cite{mbas,augusto,iru,abd,MBJFM}. A review can be found in
\cite{review}.

In the presence of anomalous abelian factors in the gauge group,
St\"uckelberg mixing with the axions render the ``anomalous''
gauge fields massive. The associated gauge symmetry is therefore
broken. In the heterotic string, with a single anomalous U(1), such a mass is always fixed at
the string scale \cite{dsw}. The situation in orientifold vacua is
richer and the masses depend non-trivially on volume and other
moduli, allowing the physical masses of anomalous U(1) gauge
bosons to be much smaller than the string scale, \cite{akr}.

If the anomalous U(1) gauge boson  masses are in the TeV range,
they  behave like $Z'$ gauge bosons widely studied in the
phenomenological literature \cite{Z1}-\cite{Z4}. One of the main
points of this paper is that unlike other Z's discussed in
heterotic string vacua as well as in unified models, the $Z'$s
associated to anomalous U(1)'s have other characteristic
low-energy couplings. These are cubic couplings between various
massive gauge bosons. Although their strength is of one-loop
caliber, they can differentiate between different types of $Z'$s.

An important role in our analysis is played by local gauge
non-invariant terms in the effective action that we call
generalized Chern-Simons terms (GCS), whose connection and role in
the anomaly cancellation is one of the main goals of this paper.
The presence of such couplings was first pointed out in
\cite{gcs1}, arising in the study of D-brane realizations of the
Standard Model. They have been independently discovered in various
supergravities in \cite{gcs2}-\cite{gcs5} and in higher
dimensional gauge theories \cite{gcs6,gcs7}.

In order to describe the relevant structure, we start from the
anomaly-related terms in the effective action
\ba {\cal S} & = & - \sum_i \int d^4x {1 \over 4 g_i^2}
F_{i,\m\n}  F_i^{\m\n} - {1 \over 2} \int d^4x  \sum_I ( \partial_\m a^I +
M^I_i A_{\m}^i)^2
\ , \nonumber \\
&& + {1 \over 24 \pi^2} C_{ij}^I \int a^I F^i\wedge F^j + {1 \over
24 \pi^2} E_{ij,k} \int A^i\wedge  A^j\wedge  F^k \ , \label{i1} \ea
where $A_i$ are abelian gauge fields, $a^I$ are axions with
St\"uckelberg couplings which render massive (some of) the gauge
fields and we used form language for compactness of notation in
the last line of (\ref{i1}).

This action is gauge-variant under
\be
A^{i}\to A^i+d\e^i\sp a^I\to a^I-M^I_i\e^i
\ee
This gauge-variance is tuned to cancel the anomalous variation of
the one-loop effective action due to the standard triangle graphs.
The contribution of the  triangle graphs is scheme dependent, (see
\cite{luis} and \cite{weinberg} as well as appendix \ref{a} for a
detailed exposition). In a natural scheme where the anomalous
variation is distributed democratically among the three vertices,
the anomaly cancellation conditions read
\be t_{ijk}\ + \ E_{ijk}+E_{ikj} \ + \ M^I_i \
C^I_{jk} \ = \ 0 \ . \label{i2} \ee
Here $t_{ijk} = Tr (Q_i Q_j Q_k)$ are the standard anomaly traces
and $Q_i$ is the charge generator associated to $A_i$.

The GCS terms are known to be scheme dependent. However, the
schemes that are relevant are typically model dependent, and it is
more convenient to expose this asymmetry in the GCS terms
explicitly. Moreover, there are combinations of GCS and axionic
terms that are gauge invariant:
\be
E_{IJk}\int (\pa a^I+M^I_i A^i)\wedge (\pa a^J+M^J_j A^j)\wedge F^k
\label{i3}\ee
Such gauge-invariant combinations lead to observable consequences.

An interesting related  question is the following. String theory
has been for a long time in search of a convincing, low energy
signature of its existence. Despite several hints over the years
it is fair to say that no such signature is known. The question
can be posed as follows: considering the particle physics data up
to a given energy (say LHC energies), is there a signature that
would rule out a UV completion by an asymptotically free or
asymptotically conformal QFT? Obviously we are keeping gravity out
of this question as no QFT UV completion is known. Anomalous
U(1)'s are ubiquitous in string theory, and it seems a good arena
to search for such signatures.

The types of couplings we are investigating in this paper are
related to triple gauge-boson couplings. For example, suppose
there is a CP-odd three-boson coupling $Z' \ Z \ \gamma$. This may
lead to a small but detectable experimental signal. Can a
consistent renormalizable gauge theory lead to a similar effect?
As we indicate, the answer turns out to be yes. We show this by
generalizing the work of \cite{gw,dhf}. In particular we  consider
the explicit example of a consistent chiral gauge theory that
emerges after the decoupling of chiral fermions, charged with
respect to the massive gauge fields and acquiring a large mass via
Yukawa couplings to Higgs bosons.

The three-gauge-boson anomalous couplings we discuss in this paper
have nontrivial consequences such as $Z' \rightarrow Z \ \gamma $
decays, which were not considered in the past in the context of
$Z'$ models \cite{Z1,Z3,KN}. A future detailed analysis of their
experimental consequences would be important and could distinguish
between models with standard anomaly cancellation and models with
a generalized anomaly cancellation mechanism.

We summarize our results as follows:

\begin{itemize}

\item The starting point is a detailed analysis of a low-energy
effective action (LEEA) which contains several U(1) gauge fields
and axions. Some of the U(1) fields get a mass via St\"uckelberg
couplings to  some axions while others remain massless. The axions
may be string theory RR axions or just the phases of Higgs field
that break the gauge symmetry. We also include non-abelian gauge
fields. There are two classes of gauge-non-invariant terms. The
axionic couplings $C^I_{ij}a^I F^i\wedge F^j$ as well as the GCS
terms $E_{ijk} A^i\wedge A^j\wedge F^k$. It is shown that the full
anomaly related effective action is fully fixed by:

(i) anomaly cancellation.

(ii) The anomaly related change traces $t_{ijk}=Tr[Q_iQ_jQ_k]$

(iii) The gauge invariant combinations of GCS and axionic terms in
(\ref{i3}). From now on we will call these terms the
``gauge-invariant GCS terms".

The rest of the terms are determined by anomaly cancellation and
depend on the scheme used to define the triangle contributions. We
use a universal symmetric scheme, that has the advantage of being
easy to use and model independent. It should be stressed that the
gauge-invariant GCS terms are scheme-independent.

\item We investigate in detail the structure of the
anomaly-related effective action in orientifold models based on
orbifold vacua of string theory, extending the analysis of
\cite{iru}. In particular, we carefully compute, the disk
coefficients $M^I_i$, ${C^I}_{ij}$ by a factorisation of one-loop
data. Many details of the orbifold geometry are important in order
to achieve this factorisation. In the process we explain the
general procedure.

We also compute the charge traces and verify that the GCS terms
associated with antisymmetric pieces of $M^I_i {C^I}_{jk}-M^I_j
{C^I}_{ik}$ are generically non-zero. We give the general
algorithm for their computation, and provide detailed calculations
in the $Z_6$ and $Z_6'$ orientifolds. Moreover we show that the
gauge-invariant GCS terms are generically non-zero, in string
theory, as is the case in supergravity analyses \cite{gcs4}. This
is a new result, as such terms are not fixed by anomalies.

\item We compute the three-anomalous-gauge-boson one-loop open
string amplitude and show explicitly that it is gauge invariant.
This together with the disk couplings completes the string theory
calculation of the relevant effective action.

\item We analyze a similar situation in QFT. We consider a theory
with an anomaly free set of chiral fermions, we give masses via
Yukawa couplings to an anomalous subset of them, and compute the
LEEFT at scales much smaller than the masses of heavy fermions.
This LEEFT is of a similar kind to that of anomalous U(1)'s coming
from string theory. We extend the previous computations of the
anomalous effective action in \cite{gw,dhf} to the general case,
and derive the gauge-invariant GCS terms. They are generically
non-trivial. The question of determining the UV charge spectrum
from the low energy GCS terms does not have a unique solution. In
particular an anomaly-free set of heavy fermions contributes
non-trivially to the gauge-invariant GCS terms.

\item We compute the full three-point amplitude at low energy of
three  U(1) gauge bosons. Some of them may be anomalous. Such
amplitudes, although one-loop in strength, are important in
characterizing the nature of Z' gauge bosons in colliders.

\item We find no determining characteristic at low energy that
would distinguish stringy anomalous U(1)'s from field theory
effective anomalous U(1)s. This however is not exclusive. More
analysis maybe necessary in this direction. An analogous question
involves non-abelian symmetries. There are no known anomalous
non-abelian symmetries in string theory. It is not clear whether
there can be such effective symmetries in QFT. This question
deserves further study.

\end{itemize}

The plan of our paper is as follows: In Section 2 we present a
general analysis of the anomaly-related effective action  with a
generalized anomaly cancellation mechanisms, including the GCS
terms. Section 3 explains the way the GCS couplings appear in
orientifold models, gives a general criterion for their existence
and formulae for axionic couplings and mixings.The explicit
examples of the $Z_6$ and $Z_6'$ orientifolds are analyzed in
detail in appendix C. In section 4 we compute the relevant
one-loop three gauge boson amplitude as well as the related by
supersymmetry  gauge-boson $\to$ 2 gaugini amplitude. In section
five we compute the  GCS and axionic terms emerging from
integrating out massive fermions in QFT. Section 6 contains the
calculation of triple effective gauge boson couplings.

In appendix A we review issues associated with the regularization,
and calculation of triangle diagrams as well as their scheme
dependence. Appendix B contains a collection of formulae relevant
for the diagonalization of the arbitrary gauge boson action in
described in section 2. Finally, Appendix D contains details of
the computation of the full three gauge boson amplitude discussed
in Section 6.

\section{The general low-energy anomaly-related
 effective action\label{general}}

In this section we will perform a general analysis of the terms in
the four-dimensional effective action relevant for anomaly
cancellation. We will consider  several anomalous U(1) vector
bosons, $A^i_{\mu}$, $i=1,2,\cdots, N_V$ with field strengths
\be
F^i_{\m\n}=\pa_{\m}A^i_{\n}-\pa_{\n}A^i_{\m} \ . \label{1}\ee
We also consider non-abelian gauge bosons $B_{\m}$ with
non-abelian field strengths
\be
G_{\m\n}=\pa_{\m}B_{\n}-\pa_{\n}B_{\m}+[B_{\m},B_{\n}] \ .
\label{2}\ee
In orientifold vacua, both types of gauge fields will originate in
the open sector. Finally, there will be a set of axion fields
$a^I$, $I=1,2,\cdots, N_a$. Some will originate in the RR sector
of the closed string sector while others will be the phases of
open string charged scalars.

We will first start from the effective Lagrangian describing the
kinetic terms of the fields
\bea {\cal L}_{kin}&=& - {1\over 2} \sum_{\a} f_{\a\b}{\rm
Tr}[G_{\a,\m\n}G_{\b}^{\m\n}] - {1\over 4}\sum_{i,j} f_{ij}
F_{i,\m\n}F_j^{\m\n} \nonumber \\
&&- {1\over 2} \sum_{I,J} h_{IJ} (\pa_{\mu} a^I + \sum_i M^I_i
A^i_{\mu})(\pa^{\mu} a^J+ \sum_i M^I_i A^{i\mu}) \label{3}\eea
We have labeled the various simple factors of the non-abelian
group\footnote{In our conventions ${\rm Tr}(t^A t^B) = 1/2
\delta^{AB}$. } with the index $\a= 1,... N_{YM}$, the Abelian
factors with the index $i=1,... N_V$ and  the axions with the
index $I= 1,... N_a$. From now on we will assume the summation
convention: repeated indices are always summed over, unless
otherwise stated.

In principle, the kinetic functions \footnote{Gauge invariance
requires $f_{i\a}=0$.} $f_{\a\b}$, $f_{ij}$ and $h_{IJ}$ depend on
dilaton-like moduli $\varphi$.  The dynamics of the latter is
irrelevant for our present purposes and we can assume they are
frozen at some non-singular value. At a given point in the moduli
space, linear combinations of $A^i_{\mu}$, $A^\a_{\mu}$ and $a^I$
put the kinetic terms in canonical form $f_{\a\b}=\delta_{\a\b}
(1/g_{\a}^2)$, $f_{ij}=\delta_{ij} (1/g_{i}^2)$ and
$h_{IJ}=\delta_{IJ}$. This results into a redefinition of the
mixing coefficients $M_I^i\rightarrow
\hat{M}_{\hat{I}}^{\hat{i}}$. Henceforth we assume we have
performed this step and simply drop the hats.

Because of the mixing with the axions, a subset of the U(1) gauge
bosons will eventually be massive. In string theory there are two
sources for these St\"uckelberg couplings. The first is
spontaneous symmetry breakdown ('Higgsing'), as in field theory.
In this case $M$ is proportional to the charge of the (Higgs)
scalar obtaining a vev. The associated  axion is the phase of the
(open string) Higgs scalar. The second source of mixing
('axioning'), as in higher dimensional (supergravity) theories,
emerges from the disk couplings between anomalous (open string)
U(1) gauge fields and axions in the RR sector of the closed-string
spectrum.

It is important for our subsequent purposes to separate the
massive from the massless U(1) gauge fields. To implement this, we
will diagonalize the mass matrix of the gauge bosons:
\be M_{ij}^2\equiv M^I_{i}M^{I}_{j} \ . \label{4}\ee
In particular we will be careful to separate the zero eigenvalues.
We will label by letters $m,n,\cdots$ the eigenvectors with zero
eigenvalue, and with $a,b,\cdots$ the eigenvectors with non-zero
eigenvalue.
\be M^2_{ij}~\eta^a_{j}=M_a^2 ~\eta^a_{i}\sp
a=1,2,\cdots N_{\bullet}\sp M_a\not=0~~\forall ~a \label{5} \ , \ee
\be
 M^2_{ij}~\eta^m_{j}=0\sp m=1,2,\cdots N_{\rm o}\sp N_{\rm o}+N_{\bullet}=N_V
\ . \label{6}\ee
The eigenvectors can be chosen to satisfy the orthonormality
conditions
\be
\eta^a_i~\eta_i^b=\delta_{ab}\sp\eta^m_i~\eta_i^{n}=\delta_{mn}\sp
\eta^m_i~\eta_i^{a}=0 \sp M^I_i~\eta_i^m=0 \ . \label{7}\ee
We also define the $\bo$ vectors in the space of axions
\be
W^I_a={M^I_i\eta_i^a\over M_a} \label{8} \ . \ee
This set is orthonormal using (\ref{7}), (\ref{8})
\be
W^I_a~W^I_b={M^I_i\eta_i^a\over M_a}{M^I_j\eta_j^b\over M_b}{M^I_iM^I_j\eta_i^a\eta_j^b\over M_aM_b}=\delta_{ab} \ .
\label{9}\ee
In general $N_{a}, N_V\geq \bn$ so we may complete (\ref{9}) into
a full basis in axion space by introducing
\be W^I_u~ W^I_v=\delta_{uv}\sp
u,v=1,2\cdots, N_{\rm inv}=N_{a}-\bn\sp   W^I_u ~W^I_a=0 \ .
\label{10}\ee

We may use now the various vectors to define new fields as follows
\be A_i= \eta^a_i ~Q^a+\eta^m_i ~Y^m\sp a^I= W^I_u ~ b^u+M_a W^I_a
~b^a \ . \label{11}\ee
The kinetic terms (\ref{3}) in the new basis
read
\bea
{\cal L}_{kin}&=&-{1\over 2}{\rm Tr}[G_{\a,\m\n}G_{\a}^{\m\n}]-
{1\over 4}F_{a,\m\n} F^{\m\n}_a-{1\over 4}F_{n,\m\n} F^{\m\n}_n\nn\\
&&-{1\over 2}\pa_\mu b^u\pa^\mu b^u -{1\over 2} M_a^2(\partial
b^a+Q^a)^2 \ . \label{12}\eea
Therefore, $Q^a_{\m}$ denotes the $\bn$ massive U(1) gauge fields,
$b^a$ their associated St\"uckelberg fields, $Y^m_{\m}$ the $\bo$
massless U(1) gauge fields, and $b^u$ the $N_{\rm inv}$ gauge
invariant axions.

The relevant infinitesimal U(1) gauge transformations are
\be
Q^a_{\m}\to Q^a_{\m}+\pa_{\m}\ve^a\sp b^a\to b^a-\ve^a\sp
Y^m_{\m}\to Y^m_{\m}+\pa_{\m}\ve^m \ , \label{13}\ee
while the non-abelian ones read
\be B_{\m}\to B_{\m}+D_{\m}\ve\sp
D_{\m}\ve\equiv \pa_{\m}\ve+[B_{\m},\ve]\sp
 G_{\m\n}\to G_{\m\n}+[G_{\m\n},\ve] \ .
\label{14}\ee
Under the above gauge transformations the kinetic terms are
obviously invariant.

We will now introduce the classically gauge non-invariant terms of
the effective action. Their ultimate goal will be to cancel the
potential one-loop triangle anomalies. They are of two types. The
first involves the Peccei-Quinn terms
\bea
{\cal L}_{PQ}&=& {b^u\over 24\pi^2}({C^u}_{ab}F^a\wedge F^b+{C^u}_{am}F^a\wedge F^m+
{C^u}_{mn}F^m\wedge F^n+{D^u}_{\a}Tr[G_{\a}\wedge G_{\a}])\nn\\
&+& {b^a\over 24\pi^2}({C^a}_{bc}F^b\wedge F^c+{C^a}_{bm}F^b\wedge
F^m+{C^a}_{mn} F^m\wedge F^n+{D^a}_{\a}Tr[G_{\a}\wedge G_{\a}]).~~~~~~ \label{15}\eea
${\cal L}_{PQ}$ contains all possible Peccei-Quinn terms. We have
used the form notation where
\be F={1\over
2}F_{\m\n}~dx^{\m}dx^{\n}\sp F\wedge F={1\over 4}
F_{\m\n}F_{\r\s}~dx^{\m}dx^{\n}dx^{\r}dx^{\s} \ . \label{16}\ee

Under gauge transformations (\ref{13}) and (\ref{14}) the
Peccei-Quinn terms transform as
\be \delta {\cal L}_{PQ}=-{\e_a\over 24\pi^2}({C^a}_{bc}F^b\wedge
F^c+{C^a}_{bm}F^b\wedge F^m+{C^a}_{mn} F^m\wedge
F^n+{D^a}_{\a}Tr[G_{\a}\wedge G_{\a}]) \label{17}\ee

The second set of gauge variant terms are the generalized
Chern-Simons terms (or GCS terms for short). They are obtained by
contracting the dual of the CS form with a gauge field. In the
abelian case we may therefore write
\be {\cal S}^{ijk}\equiv {1\over
48\pi^2}\int \epsilon^{\mu\nu\rho\sigma}
A^i_{\mu}A^{j}_{\nu}F^k_{\rho\sigma}\sp {\cal S}^{ijk}=-{\cal
S}^{jik} \ . \label{18}\ee
Under U(1) gauge transformations, $A^i\to
A^i+d\e^i$
\be \delta {\cal S}^{ijk} = {1\over 24\pi^2} \int
(\epsilon^j~F^i\wedge F^k-\epsilon^i~F^j\wedge F^k) \ .
\label{19}\ee
Not all abelian GCS are independent. We have
\be {\cal S}^{ijk}+{\cal S}^{kij}+{\cal S}^{jki} = {1\over 48\pi^2}\int
\e^{\m\n\r\s}\pa_{\m}(A^i_{\n}A^j_{\r}A^k_{\s})=0 \ . \label{20}\ee
This relation indicates that when $i=k$ or $j=k$, there is a
single independent GCS term. When all three indices are distinct,
then there are two independent GCS terms.

To define the analogous GCS terms involving the non-abelian fields
we introduce the standard non-abelian CS form
\be
\Omega_{\m\n\r}={1\over 3}{\rm Tr}\left[B_{\m}(G_{\n\rho}-{1\over
3}[B_{\n},B_{\rho}])+cyclic\right] \sp {1\over
2}\pa_{\m}\Omega_{\n\r\s}~dx^{\m}dx^{\n}dx^{\r}dx^{\s}={\rm
Tr}[G\wedge G] \label{21}\ee
which under infinitesimal gauge transformations transforms as
\be \delta\Omega_{\m\n\r}={1\over
3}Tr\left[\pa_{\m}\e
(\pa_{\n}B_{\rho}-\pa_{\rho}B_{\n})+cyclic\right] \ . \label{22}\ee
Using the CS 3-form we may now construct the mixed GCS terms
\be
{\cal S}^{i,\a}={1\over 48\pi^2}\int
\e^{\m\n\r\s}A^i_{\m}\Omega^{\a}_{\n\r\s} \ . \label{23}\ee
Under infinitesimal abelian and non-abelian gauge transformations
it transforms as
\be \delta {\cal S}^{i,\a}={1\over 24\pi^2}\int
F^i\wedge {\rm Tr}[\ve~ \tilde G_{\a}]-\ve^i{\rm Tr}[G_{\a}\wedge
G_{\a}] \ , \label{24}\ee
where $\tilde G_{\a}$ is the abelian part of $G_{\a}$. The most
general  set of GCS terms (irreducible under (\ref{20}) is given
by
\bea {\cal L}_{GCS}={1\over
48\pi^2}\e^{\m\n\rho\s}\left[E_{mnr}~Y^m_{\m}Y^n_{\nu}F^r_{\rho\s}
+E_{man}~Y^m_{\m}Q^a_{\nu}F^n_{\rho\s}+E_{mab}~Y^m_{\m}Q^a_{\nu}
F^b_{\rho\s}\right.~~~\nn\\
\left.+E_{abc}~Q^a_{\m}Q^b_{\nu}F^c_{\rho\s}+
({Z^m}_{\a}~Y^m_{\m}+{Z^a}_{\a}~Q^a_{\m})\Omega^{\a}_{\n\r\s}\right]
\ .\label{25}\eea
The coefficients satisfy the following symmetry properties
\be E_{mnr}=-E_{nmr}\sp E_{abc}=-E_{bac} \ . \label{26}\ee
The variation under infinitesimal gauge transformations takes the
form
\bea \int \delta{\cal L}_{GCS}&=&{1\over 24\pi^2}\int
E_{mnr}(\ve^n F^m\wedge F^r-\ve^m F^n\wedge F^r) +E_{man}(\ve^a
F^m\wedge F^n-\ve^m F^a\wedge F^n)\nn\\
&&+E_{mab}(\ve^a F^m\wedge F^b-\ve^m F^a\wedge F^b)
+E_{abc}~(\ve^b F^a\wedge F^c-\ve^a F^b\wedge F^c) \nn\\
&&+ {Z^m}_{\a}(F^m\wedge {\rm Tr}[\ve \tilde G^{\a}]-\ve^m{\rm
Tr}[G^{\a}\wedge G^{\a}])+ {Z^a}_{\a}(F^a\wedge {\rm Tr}[\ve
\tilde G^{\a}]-\ve^a{\rm Tr}[G^{\a}\wedge G^{\a}])\nn\\\nn\\
&=&{1\over 24\pi^2}\int \ve^m\left[-2E_{mnr}~F^n\wedge
F^r-E_{man}~F^a\wedge F^n-E_{mab}F^a\wedge F^b- {Z^m}_{\a}{\rm
Tr}[G^{\a}\wedge G^{\a}]\right]\nn\\
&&+\ve^a\left[-2E_{abc}~F^b\wedge F^c+E_{man}~F^m\wedge
F^n+E_{mab}F^m\wedge F^b-{Z^a}_{\a}{\rm Tr}[G^{\a}\wedge
G^{\a}]\right] \nn\\
&&+ \ ({Z^m}_{\a}~F^m+{Z^a}_{\a}F^a)\wedge {\rm Tr}[\ve \tilde
G^{\a}] \ . \label{27}\eea

We may now consider the non-invariance of the effective action due
to the anomalous triangle graphs. This is described in detail in
appendix \ref{a}. We use the totally symmetric scheme of defining
the triangle graphs. We obtain the anomalous gauge variation
\bea \int \delta{\cal L}_{\rm triangle}&=& -{1\over 24\pi^2}\int
\left[t_{abc}\ve^a ~F^b\wedge F^c+t_{mnr}\ve^m ~F^n\wedge F^r
\right. \nn\\
&&+t_{mab}(2\ve^a ~F^b\wedge F^m+\ve^m ~F^a\wedge F^b)+
t_{amn}(2\ve^m ~F^a\wedge F^n+\ve^a ~F^m\wedge F^n)\nn\\
&&+{T^{a}}_{\a}(2{\rm Tr}[\ve~\tilde G^{\a}]\wedge F^a+\ve^a
~{\rm Tr}[G^{\a}\wedge G^{\a}]) \nn\\
&& +\left.{T^m}_{\a}(2{\rm Tr}[\ve~\tilde G^{\a}]\wedge F^m+\ve^m
~{\rm Tr}[G^{\a}\wedge G^{\a}])
\right]\nn\\\nn\\
&=&-{1\over 24\pi^2}\int \ve^a\left[t_{abc}~F^b\wedge
F^c+2t_{mab}~F^b\wedge F^m+ t_{amn}~F^m\wedge
F^n+{T^{a}}_{\a}~{\rm Tr}[G^{\a}\wedge G^{\a}]\right]\nn\\
&&+\ve^m\left[t_{mnr}~F^n\wedge F^r+t_{mab}~F^a\wedge F^b+
2t_{amn}~F^a\wedge F^n+{T^m}_{\a}~{\rm Tr}[G^{\a}\wedge
G^{\a}]\right]\nn\\
&&+2{T^{a}}_{\a}~{\rm Tr}[\ve~\tilde G^{\a}]\wedge
F^a+2{T^m}_{\a}~{\rm Tr}[\ve~\tilde G^{\a}]\wedge F^m \ .
\label{30}\eea
The tensors $t$ and $T$ are given by the cubic traces of the
U(1) and non-abelian generators, ${\cal Q}_a$, ${\cal Q}_m$,
$T$,
\bea t_{abc}={\rm Tr}[{\cal Q}_a{\cal Q}_b{\cal Q}_c]\sp
t_{mab}={\rm Tr}[{\cal Q}_a{\cal Q}_b{\cal Q}_m]\sp t_{amn}={\rm
Tr}[{\cal Q}_a{\cal Q}_m{\cal Q}_n] \ , \nn\\
t_{mnr}={\rm Tr}[{\cal Q}_m{\cal Q}_n{\cal Q}_r] \sp
{T^{a}}_{\a}={\rm Tr}[{\cal Q}_a(TT)^{\a}]\sp {T^m}_{\a}={\rm
Tr}[{\cal Q}_m(TT)^{\a}] \ ,\label{31}\eea
with $(TT)^{\a}$ the quadratic Casimir of the $\a$-th non-abelian
factor. We have also assumed that the non-abelian cubic anomaly
cancels.

Demanding gauge invariance of the total Lagrangian
\be {\cal L}={\cal L}_{k}+{\cal L}_{PQ}+{\cal L}_{GCS}+{\cal
L}_{\rm triangle} \ , \label{32}\ee
we obtain the following conditions
\bea E_{abc}+E_{acb}+{C^a}_{bc}+t_{abc}=0 \ , \label{33}\\
-E_{mab}+{C^a}_{bm}+2t_{mab}=0 \ , \label{34}\\
-{1\over 2}(E_{man}+E_{nam})+{C^a}_{mn}+t_{amn}=0 \ ,
\label{35}\\
{Z^a}_{\a}+{D^a}_{\a}+{T^{a}}_{\a}=0 \ , \label{36}\\
t_{mnr}+2E_{mnr}=0 \ ,\label{37}\\
t_{mab}+{1\over 2}(E_{mab}+E_{mba})=0 \ , \label{38} \\
2t_{amn}+E_{man}=0 \ , \label{39}\\
{Z^m}_{\a}+{T^m}_{\a}=0 \ , \label{40}\\
2{T^a}_{\a}-{Z^a}_{\a}=0\sp 2{T^m}_{\a}-{Z^m}_{\a}=0 \ .
\label{41}\eea
Conditions (\ref{33})-(\ref{36}) stem from the invariance under
broken (massive) gauge transformations. Conditions
(\ref{37})-(\ref{40}) stem from the invariance under unbroken
(massless) gauge transformations. Finally conditions (\ref{41})
stem from nonabelian gauge invariance.

We now proceed to investigate some immediate implications of the
invariance conditions above. (\ref{40}) and (\ref{41}) imply that
\be {Z^m}_{\a}={T^m}_{\a}=0 \ , \label{42}\ee
that is, the mixed abelian/non-abelian anomaly of the massless
U(1)'s vanishes. This is indeed the case in all known orientifold
examples. In (\ref{37}) the tensor $t_{mnr}$ is completely
symmetric while $E_{mnr}$ is antisymmetric in the first two
indices. Therefore, this equation is only consistent if
\be t_{mnr}=E_{mnr}=0 \ . \label{43}\ee
This implies that the massless U(1)'s should have no cubic anomaly
among themselves. This is indeed the case in all known orientifold
examples.

Solving (\ref{34}), (\ref{35}), (\ref{38}), (\ref{39}) we obtain
\be
E_{mab}+E_{mba}=-2t_{mab}\sp {C^a}_{bm}=-3t_{mab}+{1\over
2}(E_{mab}-E_{mba}) \label{44}\ee \be \sp E_{man}={2\over
3}{C^a}_{mn}=-2t_{amn} \ . \ee
Solving (\ref{36}) and (\ref{41}) we obtain
\be {Z^a}_{\a}=-{2\over 3}{D^a}_{\a}=2{T^a}_{\a} \ .
\label{45}\ee

A counting of parameters in the anomaly equations is in order in
order to motivate the general solution given below.

In equation (\ref{33}), $t$ has the symmetry $\Yboxdim8pt\yng(3)$
and therefore ${\bn(\bn+1)(\bn+2)\over 3!}$ independent
components. In appendix \ref{a} we show that the tensor $E$ has
the symmetry $\Yboxdim8pt\yng(2,1)$ and therefore
${\bn(\bn^2-1)\over 3}$ independent components. $C$ has the
structure $\Yboxdim8pt\yng(2)\otimes \Yboxdim8pt\yng(1)$ and
therefore ${\bn^2(\bn+1)\over 2}$ components. We have
\be \Yboxdim8pt\yng(2)\otimes
\Yboxdim8pt\yng(1)=\Yboxdim8pt\yng(3)\oplus \Yboxdim8pt\yng(2,1)
\label{46}\ee
Eqs (\ref{33}) is a set of ${\bn^2(\bn+1)\over 2}$ independent
equations.

In equations (\ref{34}) and (\ref{38}), $t$ has
${\bo\bn(\bn+1)\over 2}$ independent components, while $E$ and $C$
have $\bo\bn^2$ each. The number of independent equations is
$\bo\bn^2$ for  (\ref{34}) and   ${\bo\bn(\bn+1)\over 2}^2$ for
(\ref{38}).

In equations (\ref{35}) and (\ref{39}), $t$ and $C$ have
${\bn\bo(\bo+1)\over 2}$ independent components, while $E$  has
$\bn\bo^2$. The number of independent equations is
${\bo\bn(\bn+1)\over 2}$ for (\ref{35}) and $\bo\bn^2$ for
(\ref{39}).

Finally, in equations (\ref{36}) and the first of (\ref{41}) all
tensors have $\bn N_n$ components, where $N_n$ is the number of
non-abelian group factors. This happens to also be the number of
equations.

Equations (\ref{33})-(\ref{41}) do not have a unique solution once
the charges traces are fixed. The reason is the existence of the
gauge invariant terms
\be {\cal L}_{inv}={1\over
2}\e^{\m\n\r\s}(Q^a_{\m}+\pa_{\m}b^a)(Q^b_{\n}+\pa_{\n}b^b)
\left[{\cal E}_{abc}F^c_{\r\s}+{\cal E}_{mab}F^m_{\r\s}\right] \ ,
\label{47}\ee
with ${\cal E}_{abc}=-{\cal E}_{bac}$, ${\cal E}_{mab}=-{\cal
E}_{mba}$. ${\cal E}_{abc}$ has ${\bn(\bn^2-1)\over 3}$
independent components while ${\cal E}_{mbc}$,
${\bo\bn(\bn-1)\over 2}$.

By integrating by parts, we may reabsorb the various terms in
(\ref{47}) into ${\cal L}_{PQ}$ and ${\cal L}_{GCS}$. In
particular, addition of ${\cal L}_{inv}$ to the effective action
implies the following changes in ${\cal L}$
\be {C^a}_{bc}\to {C^a}_{bc}-{\cal
E}_{abc}-{\cal E}_{acb}\sp E_{abc}\to E_{abc}+{\cal E}_{abc} \ ,
\label{49}\ee \be {C^a}_{bm}\to {C^a}_{bm}-2{\cal E}_{mab}\sp
E_{mab}\to E_{mab}-2{\cal E}_{mab} \ . \label{49a}\ee
It is obvious from (\ref{33}), (\ref{34}) and (\ref{38}) that such
shifts leave the anomaly cancellation equations invariant. We
should also remember that the PQ terms of the gauge-invariant
axions are also gauge invariant. We may use this invariance to
give the general solution to the anomaly cancellation equations
(\ref{33})-(\ref{41}).

Indeed, the general solution $(E,C,Z,D)$ to the anomaly
cancellation conditions can be written in terms of the charge
trace tensors $t_{abc}$, $t_{mab}$, $t_{amn}$, ${T^a}_{\a}$, two
arbitrary tensors ${\cal E}_{abc}$, ${\cal E}_{mab}$ satisfying
${\cal E}_{abc}=-{\cal E}_{bac}$, ${\cal E}_{mab}=-{\cal E}_{mba}$
as well as the PQ coefficients ${C^M}_{ab}$, ${C^M}_{am}$,
${C^M}_{mn}$ and ${D^M}_{\a}$. The general solution is
\be E_{abc}={\cal E}_{abc}\sp
E_{mab}=-t_{mab}+{\cal E}_{mab}\sp E_{man}=-2t_{amn} \ ,
\label{50}\ee
\be {C^a}_{bc}=-t_{abc}-{\cal E}_{abc}-{\cal
E}_{acb}\sp {C^a}_{bm}=-3t_{mab}-{\cal E}_{mab} \sp
{C^a}_{mn}=-3t_{amn} \label{51}\ee
\be {Z^a}_{\a}=2{T^a}_{\a}\sp
{D^a}_{\a}=-3{T^a}_{\a} \ . \label{52}\ee
The counting of parameters that we presented above guarantees that
this is the general solution.

The charge traces are computable from the classical action.
Therefore, to fix the full low energy action, the ${\cal E}$
coefficients, undetermined from anomaly considerations must be
calculated.

In orientifolds, this can be done by a disk calculation. To start
with, the mixing coefficients $M^I_i$, which determine which
U(1)'s become massive, are given by a disk two point function
involving an open string vector and a closed string axion.
Moreover, a disk three-point function, between two open-string
vectors and a closed string axion determines the Peccei-Quinn $C$
coefficients completely.\footnote{There are subtleties with the
normalization of the disk two and three-point functions, but they
can be eventually resolved, eg by factorizing a non planar
one-loop amplitude, to obtain an unambiguous answer.} Once the
$C$'s have been determined, the unknown gauge invariant tensors
can be evaluated as
\be {\cal E}_{abc}={1\over
4}({C^b}_{ac}-{C^a}_{bc})\sp {\cal E}_{mab}={1\over
4}({C^b}_{am}-{C^a}_{bm}) \ . \label{53}\ee
We will do this in the next section for several orientifolds and
show that the ${\cal E}$ tensors are generically non-zero. It
should be noted that even in a theory free of four-dimensional
anomalies (all cubic charge traces are zero) the gauge invariant
GCS terms may be non-zero\footnote{This may arise in a theory
where a set of anomaly-free fermions has become massive due to the
Higgs mechanism, and has been integrated out.}. This is indeed the
case in the theories of reference \cite{gcs4}.

We will write here the general solution using the original
arbitrary basis and the formulae of appendix \ref{b}. In this
basis the anomaly cancelling action is
\bea &&{\cal L}={\cal L}_{PQ}+{\cal L}_{GCS}\\
&&{\cal L}_{PQ}={{C^I}_{ij}\over 24\pi^2}~a^I~ F^i\wedge F^j
+{{D^I}_{\a}\over 24\pi^2}~a^I~{\rm Tr}[G_{\a}\wedge G_{\a}]
\label{54}\\
&& {\cal L}_{GCS}={1\over
48\pi^2}\epsilon^{\m\n\r\s}\left[E_{ijk}~
A^i_{\m}A^j_{\n}F^k_{\r\s}+{Z^i}_{\a}~
A^i_{\m}\Omega^{\a}_{\n\r\s}\right] \ . \label{55}\eea
Using
\be
Q^a=\eta^a_iA_i\sp Y^m=\eta^m_iA_i\sp b^u=W^I_u a^I\sp
b^a={W^I_a\over M_a}a^I \ , \ee
we find
\bea E_{ijk}&=&\left[-(\tilde G^{ii'}G^{jj'}-G^{ii'}\tilde
G^{jj'})\tilde G^{kk'} -{1\over 2}(\tilde G^{ii'}G^{jj'}-
G^{ii'}\tilde G^{jj'})G^{kk'}\right]t_{i'j'k'}\nn\\
&&+(\eta^a_{i}\eta^m_j-\eta^m_i\eta^a_j)\eta^b_k {\cal
E}_{mab}+\eta^a_{i}\eta^b_j\eta^c_k{\cal E}_{abc} \ , \label{56}\\
{C^I}_{ij}&=&W^I_{u}\left[{C^u}_{ab}\eta^a_i\eta^b_j+{1\over
2}{C^u}_{am}(\eta^a_i\eta^m_j+\eta^a_j\eta^m_i)
+{C^u}_{mn}\eta^m_i\eta^n_j\right]\nn\\
&&+{W^I_{a}\over M_a}\left[{C^a}_{bc}\eta^b_i\eta^c_j+{1\over
2}{C^a}_{bm}(\eta^b_i\eta^m_j+\eta^b_j\eta^m_i)
+{C^a}_{mn}\eta^m_i\eta^n_j\right]\nn\\
&=& -{M^I_k\tilde M_{kk'}t_{i'j'k'}}(G^{ii'}G^{jj'}+{3\over
2}(\tilde G^{ii'}G^{jj'}+ G^{ii'}\tilde G^{jj'}) + 3\tilde
G^{ii'}\tilde G^{jj'})
-2{M^I_k\eta^a_k\eta^b_i\eta^c_j\over M_a^2}{\cal E}_{abc}\nn\\
&&-2{M^I_k\eta^a_k\eta^b_i\eta^m_j\over M_a^2}{\cal E}_{mab}
+W^I_u\left[{C^u}_{bc}\eta^b_i\eta^c_j+
{C^u}_{mb}\eta^m_i\eta^b_j+{C^u}_{mn}\eta^m_i\eta^n_j\right] \ ,
\label{57}\eea
\bea {D^I}_{\a}&=&{D^u}_{\a}W^I_u+{D^a}_{\a}{M^I_i\eta^a_i\over
M_a^2}
\ , \label{58}\\
{Z^i}_{\a}&=&{Z^m}_{\a}\eta^m_i+{Z^a}_{\a}\eta^a_i \ ,
\label{59}\eea
where $\tilde M_{kk'}$ was defined in (\ref{b17}).

It follows from (\ref{57}) that
\bea M^I_i
{C^I}_{jk}&=&-t_{i'j'k'}\left[G^{ii'}G^{jj'}G^{kk'}+{3\over
2}(G^{ii'}G^{jj'}\tilde G^{kk'}+ G^{ii'}\tilde G^{jj'}G^{kk'})+
3G^{ii'}\tilde G^{jj'}\tilde G^{kk'}\right]\nn\\
&&-2\eta^a_i\eta^b_j\eta^c_k{\cal
E}_{abc}-\eta^a_i(\eta^b_j\eta^m_k+\eta^b_k\eta^m_j){\cal E}_{mab}
\ . \label{60}\eea
Using $\tilde G^{ii'}\tilde G^{jj'}\tilde
G^{kk'}t_{i'j'k'}=t_{mnr}\eta^m_i \eta^n_j\eta^r_k=0$, we derive
\be t_{ijk}+E_{ijk}+E_{ikj}+M^I_iC^I_{jk}=0 \ ,
\label{AnomalyCancellaton}\ee
which is the condition for gauge invariance in a generic basis.

A remark concerns the GCS terms and their relation to the scheme
dependence of the triangle anomalies. As we review in appendix
\ref{a}, all the scheme dependence of the triangle graphs is in
one to one correspondence with  the GCS terms. In particular, all
the GCS terms can be set to zero in the effective action, by
picking a particular scheme that treats the various U(1) factors
asymmetrically. However in different orientifold ground states
this scheme is vacuum dependent, as we show in the next
section. We find it more convenient to fix once and for all, the
fully symmetric scheme that treats all U(1)'s democratically and
subsequently compute, and add to the effective action the GCS
terms. This is what we  do in the sequel.

We should stress, that the important effects that we discuss in
this paper are gauge invariant and are therefore insensitive to
the choice of scheme.

We should finally stress that the scheme dependence of the GCS terms
can be also described by the non-uniqueness of the solution of the
descent equations coming from the anomaly polynomial\footnote{M.B.
and E.D. acknowledge B. Kors for a fruitful discussion on this and
related issues in connection with the results of \cite{freedkors}.
}.
Restricting ourselves for
simplicity to the abelian case, the anomaly polynomial is given by
\be I_6 \ = \ {1 \over 6} t_{ijk} F^i
F^j F^k \ = \ d I_5 \ , \label{ap1} \ee
where $t_{ijk}$ is completely symmetric. The five-form $I_5$ is only
defined up to a closed form. A solution is
\be I_5 \ = \ {1 \over 6} t_{ijk}
A^i F^j F^k \  \ , \label{ap2}
\ee
We may however add a closed form
\be
\Delta I_5=d(E_{ijk}A^iA^jF^k)
\ee
The gauge variation of $I_5$ defines the anomalies
\be \delta (I_5+\Delta I_5) \ \equiv \ d I_4 \ = \ d \ ( {1 \over
6} t_{ijk} \epsilon^i F^j F^k \ + \ {1 \over 3} E_{ijk} \epsilon^i
F^j F^k ) \ . \label{ap3} \ee
The scheme dependence is determined by the tensor $E_{ijk}$ as
advertised.

\section{Anomalies and anomalous U(1)'s in orientifold models}

It was  shown in  \cite{augusto,mbas} for 6d examples  and
in \cite{iru} for 4d orientifold vacua that
the Green-Schwarz anomaly cancellation \cite{gs} in type II and
orientifold vacua, generically involves twisted-form couplings to
gauge fields. In \cite{iru}
it was verified that indeed mixed abelian-non-abelian anomalies were
cancelled by the twisted axions. Here we will discuss the
subtleties that arise for the abelian
sector.

In the previous section we have derived  the anomaly cancellation
condition (\ref{i2}),by utilizing a symmetric scheme for the
triangle graphs. If
\be M_{\a}^I C_{\b\g}^I - M_{\b}^I
C_{\a\g}^I \ \not= \ 0 \ , \ee
then anomaly cancellation implies  the existence of generalized
Chern-Simons terms. For gauge groups coming from D-branes in type
II orientifold models, this can arise only from a non-planar
cylinder diagram that contains the (antisymmetrized) Chan-Paton
traces will be (as we will see later in (\ref{E=MC})):
\be 3 E_{\a\b\g}=
M_{\a}^I C_{\b\g}^I - M_{\b}^I C_{\a\g}^I= \sum_k \eta_k
|\sqrt{N_k}| \ tr [\gamma_k \l_{\g} \l_{[\b}] \ tr [\gamma_k
\l_{\a]} ] \ . \label{np3} \ee
%
Here $k = 1 \cdots N-1$ denotes the different type of twisted
sectors propagating in the tree-level channel cylinder diagram,
whereas
\bea N_k \ =\left\{
\begin{array}{ll}
\prod_{\Lambda=1}^3 (2 \sin [\pi k v_\Lambda])^2 &~~~~
{\rm for \ D9-D9 \ and \ D5-D5 \ sectors,} \\
\\
(2 \sin [\pi k v_3])^2 &~~~~ {\rm for \ D9-D5 \ sectors}
\end{array} \right.\label{Nk} \eea
denote the number of fixed points in the internal space and in the
third internal torus, respectively (We consider for simplicity D5
branes whose world-volume span the third internal torus $T^2_3$.).

Also, $\eta_k$ takes the values of: sign$(\prod_{\Lambda=1}^3 \sin
[\pi k v_\Lambda])$ for all sectors of D9-D9, D5-D5, D9-D5 where
the orbifold action twists all tori, $(-1)^{kv_i}$ for all sectors
of D9-D5 where the orbifold action leaves untwisted a
perpendicular torus $T^2_i$ to the D5 brane (all the above are
${\cal N}=1$ sectors), and zero for sectors of D9-D9, D5-D5, D9-D5
where the orbifold action leaves untwists the longitudinal torus
$T^2_3$ to the D5-brane (which are ${\cal N}=2$ sectors).
Notice that particles and antiparticles contribute to the anomaly
with different signs as it should be.
Let us stress that the interpretation of the factors $N_k$ is
different for D9 and D5 branes. Whereas D9 branes fill the whole
space-time and therefore couple to twisted axions localized at all
fixed points, the D5 branes can only probe some fixed points and
their associated axions. Correspondingly, their couplings to such
axions are different with respect to the D9 brane couplings.

\subsection{General formulae for the disk couplings of axions to gauge bosons}

We would like to illustrate our results in some concrete examples
such as type I compactifications on $T^6/Z_N$ orbifolds. The
resulting Chan-Paton group is typically non semi-simple and indeed
contains one (for $N=3,7$) or more abelian factors which are all
superficially anomalous. When $N$ is even, there are $Z_2$
elements ${\cal I}$ in the orbifold group.
The $\Omega
{\cal I}$ involution where $\Omega$ is the (generalized) world-sheet parity, generates
 $O5$-planes in the configuration.
D5-branes are then needed for tadpole cancellation and the gauge
group comprises two different kinds of gauge groups.

Denoting by $\gamma$ the discrete Wilson lines, projectively
embedding the orbifold group into the Chan-Paton group, we may
parameterize  them as follows:
\be \gamma^{(\alpha)}_1 = \exp(-2\pi i \oplus_r
V^{(\alpha)}_r\cdot H_r) \ee
where $H_r$ are the Cartan generators of $SO(32)^{(\alpha)}$ with
$\alpha = 9,5$. They are normalized to $tr(H_r H_s) = 2 \delta_{rs}$.
For $N$ odd, the conditions for (un)twisted RR tadpole
cancellation are
\be tr[\gamma^{(9)}_{2k}] = 32 \prod_{\Lambda=1}^3 \cos(\pi k
v_{\Lambda}) \ee
where $i$ runs over the three two-tori. Both signs in
$(\gamma^{(9)}_{1})^N =\pm 1$ are possible but the two choices
lead to equivalent physics. For even $N$ instead, only
$(\gamma^{(9)}_{1})^N = (\gamma^{(5)}_{1})^N = - 1$ is allowed.
The form of the other twisted tadpole conditions  is  model
dependent. Clearly $n_9 = n_5 =32 $ unless one turns on a
quantized NS-NS antisymmetric tensor. Moreover $\Omega^2 = 1$,
implies $(\gamma_{\Omega}^{(p)})^T = \pm \gamma_{\Omega}^{(p)}$
the standard choice is plus (+) for $p=9$ and minus ($-$) for
$p=5$.

In order to study the fate of the anomalous U(1)'s, it is
convenient to introduce the combinations
\be \lambda_i ={1\over 2\sqrt{n_i}}\sum_{r=1}^{n_i} Q^r_i H_r
\label{lamdas} \ee
where $i$ denotes the brane and $Q^r_i = (0, 0, \ldots, 0, 1,
\ldots, 1, 0, \ldots, 0)$ are 16-dimensional vectors, with $n$
one-entries at the position where the corresponding $U(n)$ lives.
Notice that $\l$'s satisfy $tr [\lambda_i \lambda_j] = {1\over 2}
\delta_{ij}$. Also
%
%
\be tr[\gamma_k \lambda_i] = -i \sqrt{n_i} \sin(2\pi k V_i) \quad
tr[\gamma_k \lambda_i \lambda_j] = {1\over 2} \cos(2 \pi k V_i)
\delta_{ij} \ . \ee
Notice that for $k=N/2$, the latter traces vanish since $V_i$ are
of the form $2 \ell +1 /N$. This sector can only contribute with
an internal volume dependent term associated to anomaly cancellation
in $D=6$.

The masses of the anomalous U(1)s have been computed in
\cite{akr}. Here we review the results:
\begin{itemize}
\item{${\cal N}=1$ Sectors:}
The contribution to the masses for ${\cal N}=1$ sectors of
$Z_N$ orbifolds, labelled by $k$, are (we assume that the D5-branes are longitudinal
to the $T^2_3$):
\begin{eqnarray}
{1\over 2}M^2_{99,k}~=~{1\over 2}M^2_{55,k}
~&=&~ - {1\over {8}\pi^3 N}~ \sqrt{N^1_k N^2_k N^3_k}~ tr[\gamma_k\lambda^a]
tr[\gamma_k\lambda^b]\label{MassesN19955}\\
{1\over 2}M^2_{95,k}
~&=&~~~~~ {\tilde{\eta}_k\over 8\pi^3 N} ~\sqrt{N^3_k}
~tr[\gamma_k\lambda^a] tr[\gamma_k\lambda^b]\label{MassesN195}
\end{eqnarray}
where, $\tilde{\eta}_k$ is ${\rm sign}\left(\prod_{\Lambda=1}^3
\sin [\pi k v_\Lambda]\right)$ when all tori are twisted and
$(-1)$ when a perpendicular torus to the D5 brane remains
untwisted by the orbifold action. Also $N_k^i=(2 \sin[\pi k
v_i])^2$ is the number of the effective fixed points of torus
$T_i^2$.

\item{${\cal N}=2$ Sectors:}
For such sectors, one $v_i k$ is integer \ie one torus is
untwisted by the orbifold action. This torus can be longitudinal
or perpendicular to the D5 branes. Without loss of generality, we
can assume that the longitudinal torus to the D5 brane is $T_3^2$
and the not untwisted-perpendicular one (if any) is $T^2_2$.

Therefore, the contribution to the masses for ${\cal N}=2$, $k$
sectors of $Z_N$ orbifolds are:
\begin{eqnarray}
{1\over 2}M^2_{99,k}~=~{1\over 2}M^2_{55,k,\parallel}~&=&~
-{2{\cal V}_3\over 4\pi^3 N}~\sqrt{N^1_k N^2_k}
~ tr[\gamma_k\lambda^a] tr[\gamma_k\lambda^b]\label{MassesN29955}\\
{1\over 2}M^2_{55,k,\perp}~&=&~ -{(2{\cal V}_{2})^{-1}\over 4\pi^3
N}~\sqrt{N^1_k N^3_k}
~ tr[\gamma_k\lambda^a] tr[\gamma_k\lambda^b]\label{MassesN255perp}\\
{1\over 2}M^2_{95,k,\parallel}~&=&~~~~~ {2 {\cal V}_3\over 4\pi^3
N} ~\tilde{\eta}_k ~ tr[\gamma_k\lambda^a]
tr[\gamma_k\lambda^b]\label{MassesN295}
\end{eqnarray}
where $\tilde{\eta}_k=(-1)^{kv_3}$ and ${\cal V}_i$ denotes the
volume of the internal torus $T^2_i$. Notice that $\parallel$ and
$\perp$ denote that the $k$th sector leaves invariant the
longitudinal (third) or a perpendicular (second) torus to the D5
brane\footnote{As an example consider the $Z_6'$ orientifold which
has vector $v=(1,-3,2)/6$. Tadpole condition implies D9 branes and
D5 branes which are longitudinal to the $T_3^2$. The $k=2,3$ are
${\cal N}=2$ sectors and the contribution to $M_{55}^2$ is given
by (\ref{MassesN255perp}), (\ref{MassesN29955}) respectively.}.
\end{itemize}

We extract the disc axionic couplings to a gauge boson $M^I_a$,
for D9-branes by factorization of the one-loop mass matrix  as
follows
\bea M^{k,f}_{a(9)}\Big|_{none} &=& {i \over \sqrt{8\pi^3 N}}
\sqrt{\ell_k^f}(N^1_k N^2_k N^3_k)^{-1/4} tr[\gamma_k \lambda_a]
\quad\quad \forall f\in {\cal F}^{123}_k \label{M9I}\\
M^{k,f}_{a(9)}\Big|_{T^2_3} ~~ &=& {i\sqrt{2 {\cal V}_3}\over
\sqrt{4\pi^3 N}} \sqrt{\ell_k^f}(N^1_k N^2_k)^{-1/4} tr[\gamma_k
\lambda_a] \quad\quad\quad~ \forall f\in {\cal F}^{12}_k
\label{M9II}\eea
where $none$, $T^2_3$ denotes the untwisted torus by the action of
the $k$th sector of the orbifold.
Notice also that we have split the sum over the index $I$
labelling the various axions into a sum over sectors labelled by
$k$ and a sum over $f$, the `effective' number of fixed points
$N_k$. $f$ spans the corresponding set ${\cal F}^{ij...}_k$.
Indices $ij...$ denote tori $T^2_i,~T^2_j,...$, where the fixed
points are placed.
D9 branes cover the entire space and pass through all fixed points.
However, they couple differently to twisted axions which are
living on these fixed points. This difference is denoted by
$\ell_k^f$, the length of the `orbit' of fixed points which are
identified under the orbifold action. In the case of geometric
orientifolds of the type $Z_N$, $\ell_k^f$ takes the values:
\bea
\ell_k^f = \left\{ \begin{array}{rccr} 1 &~&~& k \ {\rm sectors \
with \ }(k,N){\rm \ coprime } \ , \\
N/k &~&~& k ~{\rm sectors~ with} \ (k,N) {\rm \ non \ coprime \ and
} \ k < [N/2]
\end{array} \right. \
\label{Orbit}\eea
where $[N/2]$ here is the integer part of $N/2$. In the last case
in (\ref{Orbit}), we used the fact that sectors $N-k$ and $k$ are
equivalent and for $k < [N / 2]$ and all supersymmetric compact
orbifolds, $N / k$, which counts the number of fixed points
exchanged by orbifold operations, is integer for $(k,N)$ non
coprime.

For the case of D5-branes, the situation is even subtler because
D5-branes couple to a reduced number of axions \ie of fixed
points.
Here, we assume that D5-branes are longitudinal to the third torus
$T^2_3$ and they are placed at the origin of the other two tori:
\bea M^{k,f}_{a(5)}\Big|_{none} &=& {i\over \sqrt{8\pi^3 N}}
\left({N^1_k N^2_k\over N^3_k}\right)^{1/4} tr[\gamma_k \lambda_a]
\quad\quad\quad\quad~
\forall f\in {\cal F}^{123}_k \label{M5I}\\
M^{k,f}_{a(5)}\Big|_{\perp} ~~~ &=&{i(1/\sqrt{2 {\cal V}_2})\over
\sqrt{4\pi^3 N}} \left({N^1_k\over N^3_k}\right)^{1/4} tr[\gamma_k
\lambda_a] \quad\quad\quad~~
\quad \forall f\in {\cal F}^{13}_k \label{M5IIperp}\\
M^{k,f}_{a(5)}\Big|_{\parallel} ~~~~&=& {i \sqrt{2 {\cal
V}_3}\over \sqrt{4\pi^3 N}} \left({N^1_k N^2_k}\right)^{1/4}
tr[\gamma_k \lambda_a] \quad\quad\quad\quad~ \quad \forall f\in
{\cal F}^{12}_k \label{M5IIparal}\eea
where $none$, $\parallel$ and $\perp$ denote that the $k$th sector
leaves invariant none, the longitudinal or a perpendicular torus
to the D5 brane respectively.

Similarly one can extract the disk axionic couplings to two bosons
$C^I_{ab}$ as follows
\bea C^{k,f}_{ab(9)}\Big|_{none} &=& {-i\over \sqrt{2\pi^3 N}}
\sqrt{\ell_k^f}(N^1_k N^2_k N^3_k)^{-1/4} tr[\gamma_k
\lambda_a\lambda_b]
\quad\quad \forall f\in {\cal F}^{123}_k \label{C9I}\\
C^{k,f}_{ab(9)}\Big|_{T^2_3}~~ &=& {-i\sqrt{2 {\cal V}_3} \over
\sqrt{4\pi^3 N}} \sqrt{\ell_k^f}(N^1_k N^2_k)^{-1/4} tr[\gamma_k
\lambda_a\lambda_b] \quad \quad\quad~\forall f\in {\cal F}^{12}_k
\label{C9II}\eea
for D9 branes and
\bea C^{k,f}_{ab(5)}\Big|_{none} &=& {-i\over \sqrt{2\pi^3 N}}
\left({N^1_k N^2_k\over N^3_k}\right)^{1/4} tr[\gamma_k
\lambda_a\lambda_b] \quad\quad\quad\quad~
\forall f\in {\cal F}^{123}_k \label{C5I}\\
C^{k,f}_{ab(5)}\Big|_{\perp}~~~ &=& {-i(1/\sqrt{2 {\cal
V}_2})\over \sqrt{4\pi^3 N}} \left({N^1_k\over N^3_k}\right)^{1/4}
tr[\gamma_k \lambda_a\lambda_b]\quad\quad\quad\quad
\forall f\in {\cal F}^{13}_k \label{C5IIperp}\\
C^{k,f}_{ab(5)}\Big|_{\parallel} ~~~ &=& {-i \sqrt{2 {\cal
V}_3}\over \sqrt{4\pi^3 N}} \left({N^1_k N^2_k}\right)^{1/4}
tr[\gamma_k \lambda_a\lambda_b] \quad\quad\quad\quad~~~~ \forall
f\in {\cal F}^{12}_k \label{C5IIparal}\eea
for D5 branes. The normalization of $C^I$s is such that all
sectors contribute with the same footing in $M^IC^I$.
By construction, in our Chan-Paton basis there are no mixed
couplings between D5 and D9 brane anomalous U(1)'s.

\section{String derivation of anomalous couplings}

In this section we will sketch the string derivation of the
anomalous three vector boson amplitude and argue that its
anomalous variation cancels if RR tadpole cancellation takes
place. Extracting the `finite' CS terms turns out to be scheme
dependent very much as in the effective field-theory, one should
be able anyway to choose a `symmetric' scheme. We will show how
the axionic couplings can be unambiguously extracted and propose a
natural prescription for identifying the relevant regions in the
moduli space contributing to the triangle anomaly
 and to the GCS. We will also check
that spacetime supersymmetry relates the GCS couplings to
non-minimal couplings of two (neutral) `photinos' to (anomalous)
abelian vectors \cite{gcs4}.

As shown by Green and Schwarz in their seminal paper \cite{gs} and
confirmed by Inami, Kanno and Kubota \cite{ikk} in a manifestly
covariant approach, anomalous amplitudes in theories with open and
unoriented string receive contribution from the boundary of the
one-loop moduli space in the odd spin structure.   This results
from the subtle interplay between the presence of one
'supermodulus' (spin 3/2 'worldsheet gravitino' zero mode) and one
conformal Killing spinor (spin $-1/2$ zero mode) \cite{polcai}.
The former brings down the worldsheet supercurrent ${\cal G} =
\psi \cdot
\partial X$ from the action or, equivalently, requires the
insertion of $\delta(\beta)=e^{-\varphi}$ that absorbs the zero
mode of the anti-superghost $\beta= e^{\varphi}\partial\xi$. The
latter requires the insertion of $\delta(\gamma)=e^{+\varphi}$
that absorbs the zero mode of the superghost $\gamma= \eta
e^{-\varphi}$ or, equivalently, allows to fix the position in
superspace of one of the vertices, e.g. the 'longitudinal' one.

Focussing on potential gauge anomalies that only involve open string
vectors and integrating over the single fermionic supermodulus
($\bar{\chi} = \pm \chi$, depending on the reflection / boundary
conditions), one gets
\bea {\cal A}_N (k_i, \zeta_i; \zeta_N = k_N)
&=& \int_0^\infty dt \int \prod_i dy_i \int d^2 z \chi_n^{(0)} \\
&&\langle {\cal G}^n (z) \prod_i V(k_i,\zeta_i;y_i) V(k_N = -\sum_i
k_i, \zeta_N = k_N;y_N) \rangle \nonumber \eea
where $V(k_i,\zeta_i;y_i)= \z_i^\m ( \p X_\m +i k_i \cdot \y \y_\m
) e^{ik\cdot X}$ denote open string vertex operators (in the $q=0$
superghost picture) $i=1,... N-1=D/2$ (with even $D$). The last
$i=N$ `longitudinal' vertex operator $V(k_N,\zeta_N = k_N;y_N)=
k_i^\m \y_\m e^{ik\cdot X} $ can be expressed as a commutator
$V(k_N,\zeta_N = k_N;y_N)=[{\cal Q}, e^{ik\cdot X}]$. Commuting
the worldsheet supercharge ${\cal Q}$ through until one gets
\be [{\cal Q}, {\cal G}_{n}] = \gamma^m{\cal T}_{nm} \quad   \ee
and relying on the conformal Ward identity for the insertion of
the worldsheet stress tensor ${\cal T}_{nm}$ yields
\bea {\cal A}_N (k_i, \zeta_i; \zeta_N k_N) &=&{\cal I}_{ab} N_a
N_b
\int d^D X_0 d^D \psi_0 \int_0^\infty dt \times \\
&&~~~~~~{d\over dt}\prod_i dy_i\langle \prod_i V(k_i,\zeta_i;y_i)
V(k_N=-\sum_{i\neq N} k_i, \zeta_N = k_N;y_N) \rangle \nonumber \
, \eea
where $N_a$ and $N_b$ are Chan-Paton multiplicities and ${\cal
I}_{ab}$, the contribution of the sector $(a,b)$ of the internal
CFT, is a constant in the Ramond sector and coincides with the
Witten index.

Integration over (non compact) bosonic and fermionic zero modes
finally gives
\bea {\cal A}_N (k_i, \zeta_i; \zeta_N = k_N) &=&(2\pi)^{D}
\delta^D ( \sum_i k_i)
\varepsilon_{\mu_1...\mu_{D/2}\nu_1...\nu_{D/2}}
\zeta_1^{\mu_1}...\zeta_{D/2}^{\mu_{D/2}}
k_1^{\nu_1}...k_{D/2}^{\nu_{D/2}}
{\cal I}_{ab} N_a N_b \times \nonumber \\
&& \int_0^\infty dt {d\over dt} \prod_i dy_i \langle e^{ik_i X}
e^{ik_N X} \rangle \eea
Notice that most of the $t$ dependence has cancelled between
bosons and (periodic) fermions, which have the same (zero) modes
thanks to the flatness of the surface.

Irreducible chiral anomalies are associated to amplitudes such
that all vertex operators are inserted on the same boundary
\cite{morscrser}. The planar contribution from the Annulus and the
unorientable contribution from the M\"obius strip cancel against
one another after imposing RR tadpole cancellation in sectors with
non-vanishing Witten index \cite{MBJFM}.

Reducible / factorizable anomalies are associated to non planar
Annulus amplitudes such that insertions are distributed among the
two boundaries. The divergence  is regulated by momentum
flow\footnote{When one of the two boundaries accomodates only one
insertion, one needs to relax momentum conservation or to go
slightly off-shell to regulate the amplitude.} and one can extract
anomaly cancelling on shell couplings of closed string axions
(p-forms) to open string 'composites'. This is the essence of the
celebrated Green-Schwarz mechanism in $D=10$ and its
generalizations in lower dimensions
\cite{augusto,morscrser,MBJFM}. Not without some effort, in $D=4$,
where $N=3$, one can thus compute the PQ couplings $C^I_{ij}$ and
the mixing coefficients $M^I_i$, previously described. Whenever
the combination $ \sum_I C^I_{ij}M^I_k$ is not totally symmetric
(\ie generically) additional generalized Chern-Simons couplings
(GCS) are required for the gauge invariance of the EFT
description.

\subsection{Direct computation}

In principle, one could directly compute GCS in string theory in a
similar way, e.g. by relaxing $\zeta_3 = k_3 $. Thanks to the
Killing supervector in the odd spin structure, one can still fix
in superspace one of the insertions which amounts to using a
vertex $\zeta_5 \cdot \psi e^{ik_3 X}$. Integration over $\psi_0$
then yields ($13 = 6\times 5/2 -2$) terms of different kinds
depending on the choice of 4 out of 6 fermions, one from the
supercurrent, one from the 'exotic' vertex and two each from the
standard vertices,  to soak up the 4 zero modes.

We have to evaluate:
\bea
&&\int_0^\infty {dt\over t}\int_0^t {dy_1}\int_0^{y_1}{dy_2}\int{d^2z}~\times\\
&&\la
\z^1_\m(\partial X_1^\m+i k_1 \y_1 \y_1^\m)e^{ik_1 X_1}~
\z^2_\n(\partial X_2^\n+i k_2 \y_2 \y_2^\n)e^{ik_2 X_2}~
\z^3_\r \y_3^\r e^{ik_3 X_3}~
(\y_{z,\l}\partial X_z^\l+\bar{\y}_{z,\l}\bar{\partial} X_z^\l)
\ra \nonumber
\eea
Since the internal CFT contributes `topologically' to anomalous
amplitudes, the relevant contractions involve  only the non
compact bosonic coordinates and their fermion partners in the odd
spin structure. On the covering torus ${\cal T}$ the propagators
are given by:
\bea &&G(z,w)=\langle X(z,\bar{z})X(w,\bar{w})\rangle_{\cal T}={\a'\over 2}
\left(-\log\left| {\vartheta_1(z-w|\t)\over
\vartheta_1'(0|\t)}\right|^2 +2\pi {Im^2[z-w]\over Im[\t]}\right)\nn\\
&&S(z,w)=\langle \y(z,\bar{z})\y(w,\bar{w})\rangle_{\cal T} =-\p_z
G(z,w) \ . \eea The latter is bi-periodic, but not analytic. For
the Annulus with the involution $z\to\tilde{z}=1-\bar{z}$, one
gets:
\bea \langle X(z)X(w)\rangle_{\cal A}&=&
{1\over 2}\Big(G(z,w)+G(z,\tilde{w})
+G(\tilde{z},w)+G(\tilde{z},\tilde{w})\Big)= G(z,w)+G(z,\tilde{w})\nn\\
\eea
since $G(z,w)=G(\tilde{z},\tilde{w})$.

We have to evaluate two kinds of terms. Terms with 4 worldsheet
fermions and terms with 6 worldsheet fermions. The former (4
fermions terms) yield
\bea
&&\la
\z^1_\m(\p X_1^\m) e^{ik_1 X_1}~
\z^2_\n(i k_{2,\s} \y_2^\s \y_2^\n) e^{ik_2 X_2}~
\z^3_\r \y_3^\r e^{ik_3 X_3}
\y_{z,\l}\partial X_z^\l\ra =%
\nn\\&&
~~~~~~~~i~\e^{\s\n \r \l}
\z^1_\m ~p^2_\s~
\z^2_\n ~\z^3_\r~~e^{-k_1k_2
 G(1,2)}~e^{-k_1k_3 G(1,3)}~e^{-k_2k_3  G(2,3)}\times A \nn\\
&&~~~~~~~~~~\times
 \Big[i k_2^\m \partial_1 G(1,2) +  i k_3^\m \partial_1
G(1,3)\Big] \nn\\
&&~~~~~~~~~~\times\Big( \left[i k_1^\l \partial_z G(z,1)+ i k_2^\l
\partial_z G(z,2)+ i k_3^\l \partial_z G(z,3)\right]
%
+\h^{\m\l} \partial_z \partial_1 G(z,1)
\Big)~,~~~ \eea
where $A=t^{-2}$ comes from the normalization of the fermionic
zero modes ( $\sim t^{-1/2}$ each). A similar contribution is
obtained with the exchange of 1 and 2.

The latter (6 fermion terms) yield
\bea
&&\la
\z^1_\m(i k_{1,\ka} \y_1^\ka \y_1^\m)e^{ik_1 X_1}~
\z^2_\n(i k_{2,\s } \y_2^\s  \y_2^\n)e^{ik_2 X_2}~
\z^3_\r \y_3^\r e^{ik_3 X_3}
\y_{z,\l}\partial X_z^\l\ra=\nn\\
&&~~~~~~~~~~~~~~~~i p^1_\ka ~\z^1_\m ~i p^2_\s~ \z^2_\n
~\z^3_\r~~e^{-k_1k_2
 G(1,2)}~e^{-k_1k_3 G(1,3)}~e^{-k_2k_3  G(2,3)}\times\nn\\
&&~~~~~~~~~~~~~~~~A\Big(
S(1,2)[-\h^{\ka\s}\e^{\m\n\r\l}+\h^{\ka\n}\e^{\m\s\r\l}+
\h^{\m\s}\e^{\ka\n\r\l}-\h^{\m\n}\e^{\ka\s\r\l}]\nn\\
&&~~~~~~~~~~~~~~~~~~~+S(1,3)[-\h^{\ka\r}\e^{\m\s\n\l}+\h^{\m\r}\e^{\ka\s\n\l}]
+S(2,3)[-\h^{\s\r}\e^{\ka\m\n\l}+\h^{\n\r}\e^{\ka\m\s\l}]\nn\\
&&~~~~~~~~~~~~~~~~~~~+S(1,z)[\h^{\ka\l}\e^{\m\s\n\r}-\h^{\m\l}\e^{\ka\s\n\r}]
+S(2,z)[\h^{\s\l}\e^{\ka\m\n\r}-\h^{\n\l}\e^{\ka\m\s\r}]\nn\\
&&~~~~~~~~~~~~~~~~~~~+S(3,z)\h^{\r\l}\e^{\ka\m\s\n}\Big)
\Big( ik_\l^1 \p G(z,1) + ik_\l^2 \p G(z,2)+ ik_\l^3 \p
G(z,3)\Big) \ .~~~~~
\eea
The explicit presence of two extra powers of momenta makes
it clear that this contribution emerges from higher derivative string interactions.

In fact, the only potential low-derivative contribution correspond
to the contraction of $\partial X$ (in the worldsheet supercurrent)
with $\partial X(i)$ in one of the two standard vertices. This
yields
\bea {\cal A}(k_i, \zeta_i) &=& (2\pi)^{4} \delta^4 (\sum_i
k_i) \varepsilon_{\mu\nu\rho\sigma}\zeta^\mu_1 \zeta^\nu_2 k_3^\rho
\zeta^\sigma_3 \times \\
&& \int_0^\infty {dt \over t^3} \int d^2z \prod_i
dy_i \partial_z\partial_1 G(z, y_1;t) \prod_{i<j} \exp(- k_ik_j
G(i,j)) + (1\leftrightarrow 2) \ . \nonumber \eea

The integration over $z$ can be performed explicitly and yields a
constant \cite{antoetal}. If momentum conservation were imposed, the
subsequent integrations over $y_i$ would simply yield powers of $t$
since $k_i\cdot k_j =0$ for three on-shell (masseless) vectors.
Relaxing momentum conservation, which is tantamount to postponing
integration over the center of mass of the string $X_0$ until the
very end, regulates the amplitudes and allows one to identify the
various effective field theory contributions from the various
'corners' of the one-loop moduli space. Subtracting the residue of
the simple pole at $t=0$, that unambiguously yields the axionic
exchange (closed string IR $\approx$ open string UV) and splitting
the remaining $t$ integral into two regions $I_{CS} = (0, T)$ and
$I_{ta} = (T,\infty)$, it is easy to convince oneself that the
latter exposes the triangle anomaly (open string IR) while the
former exposes the GCS couplings that are generated by massive
off-shell closed string exchange or, equivalently, by massive open
strings circulating in the loop. As manifest in the need of
introducing a cutoff $T$, the last two contributions cannot be
separated unambiguously. Yet the total string amplitudes is clearly
$T$ independent. Any choice of $T$ is a choice of scheme very much
as in the effective field theory description.

\subsection{The susy analog: $\g\to 2\tilde{\g}$}

For supersymmetric theories, in the low energy effective
description one also has \cite{gcs4} :
\bea{\cal L}_{VFF} = E_{ij,k} \bar\lambda^j \sigma^\mu \lambda^k
A_\mu^i + h.c.\eea
with the same E's as in the GCS terms. This can be easily deduced in
superspace, since both GCS and the VFF coupling arise from
\bea E_{ij,k} \int d^4\theta V^i D^\alpha V^j W_\alpha^k +
h.c.\eea
Unfortunately this means that the corresponding one-loop VFF
amplitude is also naively divergent / ambiguous. In fact it
receives contribution both from the odd and the even spin
structure that neatly combine to reproduce the GCS amplitude, up
to obvious kinematic factors.

Our analysis shows that, at least for amplitudes with external
fermions, the correct prescription is to insert one picture
changing operator
\bea \G(z_0)=\{ Q_{BRS},\xi(z_0)\}=c\p\xi +e^{+\varphi} G +
{1\over 2}  e^{+2\varphi} (b\p \h -2 \p b \h) \ , \eea
rather than integrating over the supercurrent ${\cal G}$
insertion. The latter prescription would give unphysical branch
cuts in this case while it gave an equivalent and thus correct
result for the three vector amplitude above.

For this reason let us recall the form of the vertex operators in
the relevant superghost pictures
\bea &&V^R_{-1/2} = u^\a S_\a e^{-\varphi/2} \Sigma \ e^{ik\cdot X} \ , \nn \\
&&V^R_{+1/2}= \bar{v}^{\dot{\a}} \s^\m_{\dot{\a}\a} S^\a \p X_\m
e^{\varphi/2} \Sigma^+ \ e^{ik\cdot X} + ... = \lim_{z_0 \rightarrow z}\G(z_0) V^R_{-1/2}(z) \ , \nn \\
&&V^{NS}_{0}= \z^\m ( \p X_\m +i k \cdot \y \y_\m ) \ e^{ik\cdot X} \ , \nn \\
&&V^{NS}_{-1}= \z^\m \y_\m e^{-\varphi} \ e^{ik\cdot X} \ .
\eea
The relevant contributions for amplitudes with a low number of
insertions come from the action of the term proportional to
$e^{+\varphi} {\cal G} = e^{+\varphi} (\y^\l \p X_\l + {\cal
G}_{int}) $ in $\G(z_0)$ .
The internal spin fields that appear in the gaugino vertex operator can be bozonized as follows:
\bea \Sigma=e^{i(\f_2+\f_3+\f_4)/2} ~~,~~~
\Sigma^+=e^{-i(\f_2+\f_3+\f_4)/2} \ . \eea

In a given spin structure $\a$,
the amplitude that we are to evaluate is
\bea
A_{VFF} = \langle e^{\varphi} \y \p X (z_0)
~ u^\a(k_1) S_\a e^{-\varphi/2}\Sigma e^{ik_1X_1}
~ \bar{v}^{\dot{\a}}(k_2) C_{\dot{\a}} \Sigma^+
e^{-\varphi/2} e^{ik_2X_2}
~ \z_\m (\p X^\m +i k_3\y\y^\m)e^{ik_3X_3}
\rangle_{\a}
\nn\\\eea
Eventually one has to sum over both even and odd
spin-structures with the GSO projection $c_\a$.

The fermionic block is:
\bea \langle e^{\varphi} \y^\l e^{-\varphi/2} S_\a \Sigma
e^{-\varphi/2} C_{\dot{\a}} \Sigma^+(\p X^\mu +ik_3\y\y^\m)\rangle \eea
and the fermions can be bosonized as: $\y^\l\to (+1,0),~S_\a\to
(-1/2, -1/2),~ C_{\dot{\a}} ~ \to (-1/2,+1/2)$. Current algebra
Ward identities also yield
\bea ik_\n^3\langle \y^\l(0) S_\a(1) &&C_{\dot{\a}}(2) \psi^\n\psi^\m(3)\rangle =
ik_\n^3\Big[(\d^\m_\l \s^\n_{\a\dot{\a}} -\d^\n_\l \s^\m_{\a\dot{\a}})\p_3G(z_0-y_3)\nn\\
&&+{1\over 2} (\s^{\n\m})^{~\b}_{\a} (\s_{\l})_{\b\dot{\a}} \p_3G(y_3,y_1)
+{1\over 2} (\s_{\l})_{\a\dot{\b}} (\s^{\n\m})^{\dot{\b}}_{~\dot{\a}} \p_3G(y_3,y_2)\Big]
\nn\\\eea
Finally the internal orbifold CFT contributes
\bea \langle \Sigma(y_1) \Sigma^+(y_2)\rangle_\a
&=&\vartheta_1(y_{12})^{-3/4}\vartheta_\a({y_{12}\over 2})
\prod_I{\vartheta_\a({y_{12}\over 2}+kv_I) \over \vartheta_1(kv_I)}
\ , \eea
where the effect of the orbifold projection ({\it viz.} $k v_I$) has been taken into account.

Assembling the various pieces above, up to the kinematical factor $u \sigma^\mu v \zeta_\mu$, one gets
\bea
{\vartheta_\a(y_{12}/2) \over \vartheta_1(y_{12})}
\prod_{I=1}^3 {\vartheta_\a(y_{12}/2+k v_I) \over
\vartheta_1(k v_I)}\times
\Bigg[\Big\{ \h^{\l\n} \p_0\p_3 G(z_0,y_3) + \sum_{i=1}^3 i k_i^\l
\p_z G(z_0,y_i)~\sum_{j\neq 2}^3 ik_j^\n
\p_2 G(2,j) \Big\}\nn\\
+\Big\{
{1\over 2} (\s^{\n}_{\m})^{\b}_{~\a}(\s_{\l})_{\dot{\b}\a}
\left(\p_3G(y_3,y_1)-\p_3G(y_3,z_0)\right)
-
{1\over 2} (\s_{\l})_{\a\dot{\b}} (\bar{\s}^{\n\m})^{\dot{\b}}_{~\dot{\a}}
\left(\p_3G(y_3,y_1)-\p_3G(y_3,z_0)\right)\Big\}\nn\\
\times \sum_{i=1}^3 i k_i^\l \p_0 G(z_0,y_i)
\Bigg]
\nn\\\eea

Taking the limit $k_i\to 0$ and fixing the position of $y_1$, one has to perform the sum over the spin
structures (at fixed twist structure)
\bea
\sum_{\a=1}^4 c_\a
\int {dt\over t^3}\int dy_2dy_3
{\vartheta_\a(y_{12}/2) \over \vartheta_\a(y_{12})} \prod_{I=1}^3
{\vartheta_\a(y_{12}/2+k v_I) \over \vartheta_\a(k v_I)}\p_0\p_3
G(z_0,y_3) \ .
\eea
For $c_\a$ the coefficients of the GSO-projection, one can make use of the identity:
$${1\over 2} \sum_{\a\b=0}^1(-1)^{\a+\b+\a\b}
\prod_{i=1}^4\vartheta[^\a_\b](v_i)=- \prod_{i=1}^4
\vartheta_1(v_i') \ , $$
where $v_i'= -v_i+{1\over 2} \sum_l v_l$, and find that
$${1\over 2} \sum_\a \vartheta_\a(y_{12}/2)\prod_I \vartheta_\a(y_{12}/2+kv_I)
= -\vartheta(y_{12})\prod_I \vartheta_\a(kv_I) \ ,
$$
and therefore all $\vartheta_\a$s exactly cancel. One ends up with
an amplitude similar to the one for the insertion of three bosonic
VO's, that in the same $k_i\to 0$ limit reads
\bea
\sim~\int {dt\over t^3}\int dy_2dy_3
\p_0\p_3 G(z_0,y_3) \ .
\eea Once again extracting the susy counterpart of the GCS is
scheme dependent but one can unambiguously identify
(supersymmetrized) axion exchange with the residue of the simple
pole at $t=0$. Introducing an open string IR cutoff $T$ as above,
one can associate the contribution of the susy partners of the GCS
to the interval $t =(0,T)$ and the `massless' open string loop
with the region $t =(T, \infty)$. One has to keep in mind that
only the total sum is $T$ independent and thus unambiguous, the
individual contributions are non gauge invariant and thus
ambiguous (scheme dependent).

\section{Heavy fermions and low-energy effective actions }

So far we have analyzed in detail the structure of the anomaly-related
effective action for orientifold models.
We have seen, that apart from the generic appearance of anomalous U(1)'s, there is
a rich pattern of axionic couplings and GCS terms.
It is an interesting question if such patterns  emerge in EFTs of
UV-complete\footnote{We define a QFT to be UV-complete
if all gauge couplings are asymptotically free or asymptotically conformal.}
quantum field theories. In particular, we are interested in knowing,
whether in the anomaly sector of an EFT, we can distinguish whether
the UV completion is stringy or a UV-complete QFT.

To proceed we consider a consistent ({\it i.e.} anomaly-free) and
renormalizable gauge theory with spontaneously-broken gauge symmetry
via the Brout-Englert-Higgs mechanism\footnote{Strictly speaking
quartic scalar couplings necessary for the Higgs potential are IR
free. We will still call this a UV complete theory as the scalars
could be bound states of fermions.}. Through appropriate Yukawa
couplings, some large masses to a subset of the fermions can be
given. We denote by $\psi^{(H)}_{L,R}$ such massive chiral fermions.
Their $U(1)_i$ charges are $X^{(H)i}_{L,R}$. In the sequel, we will
generalize the \cite{gw,dhf} calculations of the effective anomaly
related couplings in the EFT, generated by the loops of  the heavy
chiral fermions.

The relevant terms in the effective
action of the heavy fermion sector of the theory are
\bea L_H \
&=& \ {\bar \psi}^{(H)}_L \left( i \gamma^{\m}
\partial_{\m} + X^{(H)i}_L \gamma^{\m} A_{\mu}^i \right)
\psi^{(H)}_L + {\bar \psi}^{(H)}_R \left( i \gamma^{\m}
\partial_{\m} + X^{(H)i}_R \gamma^{\m} A_{\mu}^i \right)
\psi^{(H)}_R
\nonumber \\
&& - \left( \lambda_I^H \phi_I   {\bar \psi}^{(H)}_L \psi^{(H)}_R
+ {\rm h.c.} \right) \ , \label{decoupling1} \eea
where  $\phi^I$
are a set of Higgs fields of $U(1)_i$ charges $X_I^i$. They spontaneously break
the abelian gauge symmetries via their vevs, $\langle \phi^I \rangle$
 We are interested in a chiral fermion set,
\be X^{(H)i}_L \ - \ X^{(H)i}_R \ = X_I^i \not= \ 0 \ .
\label{decoupling2} \ee
If the associated Yukawa couplings are
large,  $\lambda_I^H \gg g_i$, spontaneous symmetry breaking
generates large Dirac fermion masses $M_H =\lambda_I^H v_I  $,
where $\langle \phi_I \rangle = v_I$. We consider the heavy
fermion decoupling limit, with fixed Higgs vev's and fixed gauge
boson masses, whereas $M_H \rightarrow \infty$.  If the initial
theory were anomaly-free, i.e.
\be \sum_l (X_L^i X_L^j X_L^k -
X_R^i X_R^j X_R^k )^{(l)} + \sum_H (X_L^i X_L^j X_L^k - X_R^i
X_R^j X_R^k )^{(H)} \ = \ 0 \ , \label{decoupling02} \ee
where
$(l)$ denote the massless (light) fermionic spectrum, then in the
low-energy theory ({\it i.e.} at energies below the heavy fermion
masses) with the heavy fermions integrated-out, there are
Adler-Bell-Jackiw triangle anomalies coming from the light
fermions. There will also be Wess-Zumino-like couplings generated by the loops of the
heavy fermions. The resulting low-energy action, for simple gauge groups, was worked out in
specific regularization schemes in various papers starting with
\cite{dhf}. After symmetry breaking, we parameterize the scalar
fields  by
\be \phi_I \ = \ (v_I + h_I) \ e^{i \ a_I \over v_I} \
, \label{decoupling3} \ee
where $h_I$ are massive Higgs-like
fields and $a_I$ are gauge-variant phases (axions) which will play a crucial role
in the anomaly cancellation at low-energy. To be definite, we
consider a larger number of abelian gauge fields than gauge-variant axions.
The gauge transformations of gauge fields and axions are
\be \delta
A_{\mu}^i \ = \
\partial_{\mu} \epsilon^i \quad , \quad \delta a_I \ = \ v_I \ X^i_I
\ \epsilon^i \ , \label{decoupling4} \ee
where $X^i_I$ are the
$U(1)_i$ charges of $\phi_I$.

We can compute explicitly the eventual GCS terms by performing a
diagrammatic computation starting from the action
(\ref{decoupling1}). In order to do this, we start from the
corresponding three gauge boson amplitude induced by triangle
diagram loops of heavy fermions and expand in powers of external
momenta $k_i / M_H$. We work in a basis of left ($L$) and right
($R$) fermionic fields, with the fermionic propagator having the
components
\be S_{LL} (p) = S_{RR} (p) = { - \sla{p} \over
p^2-M_H^2} \quad , \quad S_{LR} (p) = { - M_H \over p^2-M_H^2} \ .
\ee
The purely left and right propagation in the triangle loop is
similar to the computation of the anomaly with massive fermions in
the loop and will only be sketched here. The corresponding
contribution to the three-point function is of the form
\footnote{See Appendix A for notations and conventions for triangle
diagrams.}
\bea \Gamma^{\nu \rho \mu}_{ijk} (p,k_1,k_2,a) \ &=& \ i
\sum_H (X_R^i X_R^j X_R^k)~tr \Big[ \ {\sla{p}-
\sla{k_1}+\sla{a}  \over  (p-k_1+a)^2 - M_H^2}  \gamma^{\nu}
{\sla{p} + \sla{a}  \over  (p+a)^2 - M_H^2} \nonumber \\
&& \times \ \gamma^{\rho} \ {\sla{p}+ \sla{k_2}+\sla{a}  \over
(p+k_2+a)^2 - M_H^2} \gamma^{\mu} {1 + \gamma_5 \over 2} \ \Big]
   \  , \label{decoupling014}
\eea
for the right-handed fermions, where $a$ is the shift vector
\cite{weinberg}, and a similar expression with obvious changes for
left-handed ones. Only the linear terms in the expansion do
correspond to GCS terms. By expanding to linear order we obtain
\be
\Gamma^{\nu \rho \mu}_{ijk} (p ,k_i ,a) \ \simeq \ \Gamma^{\nu
\rho \mu}_{ijk} (p ,0 ,0) \ + \ k_i^{\a} {\partial \over
\partial k_i^{\a}} \Gamma^{\nu \rho \mu}_{ijk} (p ,k_i ,0)|_{k_i 0} \ + \ a^{\a} {\partial \over \partial a^{\a}} \Gamma^{\nu \rho
\mu}_{ijk} (p ,0 ,a)|_{a = 0} \ , \label{decoupling015} \ee
the shift vector is parameterized as \be a_{ijk}^{\a} = A_{ijk}
k_1^{\a} + B_{ijk} k_2^{\a}\;, \ee A straightforward computation
indicates that the term in the effective action, originating from
the second term in the r.h.s. of (\ref{decoupling015}) is
proportional to \be t^{(H)}_{ijk,L-R} \int ( A^i \wedge A^j \wedge
F^k + A^i \wedge A^k \wedge F^j) \ee
 where
\be t^{(H)}_{ijk,L-R} = (X_L^i X_L^j X_L^k)^{(H)} - (X_R^i X_R^j
X_R^k)^{(H)}\;.
\ee
It vanishes identically,  since $t^{(H)}_{ijk,L-R}$ is symmetric in all
indices, whereas the GCS terms are antisymmetric in two indices.
On the other hand, the last term in the r.h.s. of
(\ref{decoupling015}) gives a surface contribution in the loop
momentum $p$. The  surface integral is evaluated to be
\be
\int d^4 p \ \partial_{\sigma} { p^2 p_{\epsilon} \over
(p^2-M_H^2)^3} \ = \ - {\pi^2 \over 4} \eta_{\sigma \epsilon}   \
.  \label{decoupling016} \ee
The contribution to the
effective action therefore is
\be S_{GCS}^{(1)} =  {1 \over 48 \pi^2} \
\sum_H t^{(H)}_{ijk,L-R} \ \int ( A_{ijk}  A^i \wedge A^k \wedge
F^j \ - \ B_{ijk} A^i \wedge A^j \wedge F^k ) \ .
\label{decoupling017} \ee
We observe from (\ref{decoupling017}) that the contribution to the
GCS terms coming from diagrams without mass insertions are zero in
the natural scheme in which the anomaly is split democratically
between the different external currents (the symmetric scheme).
%
%
%
%
%
\begin{figure}[h]\begin{center}
\begin{tabular}{lcr}
\\
\unitlength=0.4mm
\begin{fmffile}{Mass_I1}
\begin{fmfgraph*}(60,40)
\fmfpen{thick} \fmfleft{ii0,ii1,ii2} \fmfstraight
\fmffreeze
\fmftop{ii2,t1,t2,t3,oo2} \fmfbottom{ii0,b1,b2,b3,oo1}
\fmf{phantom}{ii2,t1,t2} \fmf{phantom}{ii0,b1,b2}
\fmf{phantom}{t1,v1,b1} \fmf{phantom}{t2,b2} \fmf{phantom}{t3,b3}
\fmffreeze
\fmf{photon}{ii1,v1}
\fmf{fermion,label=$R$,l.side=right}{b3,t3}
\fmf{fermion,label=$R$,l.side=right}{t3,a}
\fmf{fermion,label=$L$,l.side=right}{a,v1}
\fmfv{decor.shape=cross,decor.size=.15w}{a}
\fmf{fermion,label=$L$,l.side=right}{v1,b}
\fmf{fermion,label=$R$,l.side=right}{b,b3}
\fmfv{decor.shape=cross,decor.size=.15w}{b}

\fmf{photon}{b3,oo1} \fmf{photon}{t3,oo2}
\fmffreeze
\fmflabel{$A^\mu_i(k_3)~$}{ii1} \fmflabel{$A^\nu_j(k_1)$}{oo2}
\fmflabel{$A^\rho_k(k_2)$}{oo1}
%
%
\end{fmfgraph*}
\end{fmffile}
&~~~~~~~~~~~~~~~~~~~~~~~&
%
%
%
\unitlength=0.4mm
\begin{fmffile}{Axion_VI1}
\begin{fmfgraph*}(60,40)
\fmfpen{thick} \fmfleft{ii0,ii1,ii2} \fmfstraight
\fmffreeze
\fmftop{ii2,t1,t2,t3,oo2} \fmfbottom{ii0,b1,b2,b3,oo1}
\fmf{phantom}{ii2,t1,t2} \fmf{phantom}{ii0,b1,b2}
\fmf{phantom}{t1,v1,b1} \fmf{phantom}{t2,b2} \fmf{phantom}{t3,b3}
\fmffreeze
\fmf{dashes}{ii1,v1}
\fmf{fermion,label=$L$,l.side=right}{t3,v1}
\fmf{fermion,label=$R$,l.side=right}{b3,a}
\fmf{fermion,label=$L$,l.side=right}{a,t3}
\fmfv{decor.shape=cross,decor.size=.15w}{a}
\fmf{fermion,label=$R$,l.side=right}{v1,b3}
\fmf{photon}{b3,oo1} \fmf{photon}{t3,oo2}
\fmffreeze
\fmflabel{$\alpha_I(k_3)~$}{ii1} \fmflabel{$A^\mu_i(k_1)$}{oo2}
\fmflabel{$A^\nu_j(k_2)$}{oo1}
\end{fmfgraph*}
\end{fmffile}
\\ \\
\end{tabular}
\caption{The first diagram is one of the twelve diagrams which
contribute to the GCS terms. The second is one of the six diagrams
which contribute to the axionic couplings. Both are obtained by
integrating out heavy fermions.} \label{Massless}
\end{center}
\end{figure}
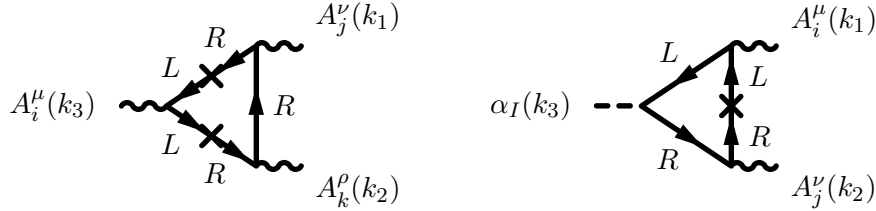

The new interesting ingredients in the massive fermion case appear
due to the mass insertions in the propagators $S_{LR} (p)$. Mass
insertions on two of the three fermionic propagators produce new
contributions which are UV finite and easily evaluated. There are
twelve new diagrams corresponding to the three possible ways of
distributing the mass insertions, to the symmetrization of the
external bosonic lines and to the two types of components (left versus
right-handed fermions) in each propagator. We portray just one
example. For mass insertions on the propagators of momenta $p - k_1$
and $p + k_2$ , one of the contributions to the three-point function
is
\bea \Gamma^{\nu \rho \mu}_{ijk} (p,k_1,k_2)^{(2)} &=&  i
\sum_H (X_L^i X_R^j X_R^k)^{(H)}  \ tr \ \Big[ \ {M_H \gamma^{\nu}\over  (p-k_1)^2 -
M_H^2}  {\sla{p} \gamma^{\rho}  \over  p^2 - M_H^2} {M_H \gamma^{\mu}
\over (p+k_2)^2 - M_H^2}~~~~
\nonumber \\
&& +   \ {M_H \over  (p-k_2)^2 - M_H^2}  \gamma^{\rho} {\sla{p}
\over  p^2 - M_H^2} \gamma^{\nu} {M_H \over  (p+k_1)^2 -
M_H^2} \gamma^{\mu}  \   {1 + \gamma_5 \over 2} ~ \Big]
\  . ~~~\label{decoupling14}
\eea
Since the result is finite, we don't need to introduce a
shift vector $a$. As before, we expand in powers of the
external momenta and we keep only the linear term. By a
straightforward computation we find
\be \int {d^4 p \over (2
\pi)^4} \Gamma^{\nu \rho \mu}_{ijk} (p,k_1,k_2)^{(2)} =
\epsilon^{\mu \nu \rho \alpha} (k_2-k_1)_{\alpha} \sum_H { 4 M_H^2
\over 3}  (X_L^i X_R^j X_R^k)^{(H)} \int {d^4 p \over (2 \pi)^4}
{1 \over (p^2-M_H^2)^3} \ . \label{decoupling15} \ee
The result
(\ref{decoupling15}) gives a finite contribution in the $M_H
\rightarrow \infty$ limit. By adding the twelve different
diagrams, we find a local term in the effective action
\be
S_{GCS}^{(2)} =  {1 \over 96 \pi^2} \sum_H (X_L^i X_R^j - X_R^i
X_L^j)^{(H)} (X_R^k+ X_L^k)^{(H)} \int A_i \wedge A_j \wedge F_k \
. \label{decoupling16} \ee
The lagrangian (\ref{decoupling1})
contains also couplings of axions to the fermions, of the type
\be
L_{\rm Yuk} \ = \ - i \ \lambda_I^H a_I  {\bar \psi}^{(H)}
\gamma_5 \psi^{(H)} + \cdots \ , \label{decoupling17} \ee
where the ellipsis stands  for higher order couplings that give no contributions in
the $M_H \rightarrow \infty$ limit. The axion-heavy fermion
couplings generate axionic couplings to gauge fields through one
mass insertion in triangle diagrams. There are six relevant
diagrams. We consider as an example the one with the mass
insertion on the fermionic propagator of momentum $p + k_2$. One
of the two contributions to the three-point function ${\tilde
\Gamma}^{\mu \nu}_{ij} (p,k_1,k_2)^{(1)}$ is equal to
\footnote{The heavy fermions H obtain their mass generically from a single Higgs, whose phase
is $a_I$. We denote this by $H_I$. Therefore, a sum over $H_I$ is over all
massive fermions who get their mass from the I-th Higgs.}
\be i \sum_{H_I}
\lambda_I^{H_I} (X_L^i X_L^j)^{(H_I)}  \  tr \ \Big[ \ \gamma_5 {\sla{p} -
\sla{k_1} \over (p-k_1)^2 - M_{H_I}^2} \gamma^{\mu} {\sla{p}  \over
p^2 - M_{H_I}^2} {M_{H_I} \over  (p+k_2)^2 - M_{H_I}^2} \gamma^{\nu} {1 -
\gamma_5 \over 2} \ \Big] \  . \label{decoupling18} \ee
It leads to
\be \int {d^4 p \over (2 \pi)^4} {\tilde \Gamma}^{\mu \nu}_{ij}
(p,k_1,k_2)^{(1)} \epsilon^{\mu \nu \alpha \beta} (k_1)_{\alpha}
(k_2)_{\beta} \sum_{H_I} \lambda_I^{H_I} M_{H_I} (X_L^i X_L^j)^{(H_I)} \int {d^4
p \over (2 \pi)^4} {p^2 \over (p^2-M_{H_I}^2)^4} \ .
\label{decoupling19} \ee
We take the limit $\lambda_I^{H_I}
\rightarrow \infty$ with fixed $v_I$. In this limit
(\ref{decoupling19}) survives and is proportional to $1 / v_I$.
Adding the five other diagrams we get the axionic couplings
\be
S_{\rm ax} = {1 \over 96 \pi^2} \sum_{I} \sum_{H_I} [ 2 (X_L^i X_L^j + X_R^i
X_R^j) +  X_L^i X_R^j + X_R^i X_L^j ]^{(H_I)} \ \int  \ {a_I \over
v_I} \ F_i \wedge F_j \ .  \label{decoupling20} \ee
On the other
hand, the kinetic terms of the Higgs fields $\phi_I$ generate
the St\"uckelberg mixings
\be |\partial_{\mu} \phi_I - i X_I^i
A_{\mu}^i \phi_I|^2 \rightarrow ( \partial_{\mu} a_I - X_I^i v_I
A_{\mu}^i)^2 \ . \label{decoupling21} \ee
We therefore find the following GCS terms, axionic couplings and
kinetic mixings
\bea
&&E_{ij,k} \ = \ {1 \over 4} \sum_H (X_L^i X_R^j - X_R^i X_L^j)^{(H)} (X_R^k+ X_L^k)^{(H)} \ , \nonumber \\
&&C^I_{ij} \ = \  {1 \over 4 v_I} \ \sum_{H_I} [ 2 (X_L^i X_L^j + X_R^i X_R^j) +  X_L^i X_R^j + X_R^i X_L^j ]^{(H_I)} \ , \nonumber \\
&&M^I_i \ = \ v_I \ X_I^i \ = \ v_I \ (X_L^i-X_R^i)^{(H_I)} \quad ,
\quad {\rm for \ every \ } H_I \ . \label{decoupling22} \eea
There are some obvious checks of the formulae above. Clearly the GCS
terms should cancel in the non-chiral case $X_L^i=X_R^i$. In the
particular chiral case $X_L^i = - X_R^i$, they should also cancel
since the GCS terms have to be antisymmetric under the left-right
interchange  $X_L^i \leftrightarrow X_R^i$ and the first term in the
GCS terms in (\ref{decoupling22}) is already antisymmetric. Notice
the following identities which prove the anomaly cancellation
conditions are satisfied as a particular case of the more general
analysis performed in Section 2
\bea && {1 \over 3} (M_i^I C_{jk}^I
+ M_j^I C_{ki}^I + M_k^I C_{ij}^I ) \ = \ {1 \over 2} \sum_H
(X_L^i X_L^j X_L^k - X_R^i X_R^j X_R^k )^{(H)} \ ,
\label{decoupling23} \\
&& {1 \over 3} (M_i^I C_{jk}^I - M_j^I
C_{ik}^I) \ = \ {1 \over 4} \sum_H (X_L^i X_R^j - X_R^i
X_L^j)^{(H)} (X_R^k+ X_L^k)^{(H)} \ = \ E_{ij,k} \ . \nonumber \eea
Moreover, in the case where all the U(1)'s are massive, comparison with (\ref{50}-\ref{52})
indicates that the gauge invariant GCS are given by the second line above.

The gauge variations of the induced GCS terms and axionic
couplings in (\ref{decoupling16}) and (\ref{decoupling20}) are
\ba
\delta ( S_{GCS} + S_{\rm ax} ) \ = {1 \over 48 \pi^2} \ \sum_H
(X_L^i X_L^j X_L^k - X_R^i X_R^j X_R^k )^{(H)} \int \epsilon_i F_j
\wedge F_k \ . \label{decoupling24} \ea
This anomalous variation
is $1/2$ compared to the standard anomaly contribution in the
appendix, (\ref{a14}). The reason is that (\ref{decoupling24}) is
not yet the full anomaly (see e.g. \cite{gcs6}). Indeed, the
classical value of the divergence of the heavy fermionic current
is
\ba
&& \partial^{\mu} J_{\mu}^{(H)} \ = \ i M_H (X_R^i-X_L^i)^{(H)} \ {\bar \psi}^{(H)} \gamma_5 \psi^{(H)} \ , \nonumber \\
&& {\rm where} \ J_{\mu}^{(H)} \ = \ X_R^{i,(H)} {\bar \psi}_R^{(H)}
\gamma_{\mu} \psi_R^{(H)} + X_L^{i,(H)} {\bar \psi}_L^{(H)}
\gamma_{\mu} \psi_L^{(H)} \ . \label{decoupling25} \ea
The matrix
element of this classical part can be evaluated diagramatically. The
computation is basically identical to the computation of the axionic
couplings above. The result, in the decoupling limit $M_H
\rightarrow \infty$, is
\ba && \langle 0 \ | \ i \ M_H \
(X_R^i-X_L^i)^{(H)} \ {\bar \psi}^{(H)} \gamma_5 \psi^{(H)} |
A_j^{\mu} (k_1) A_k^{\nu} (k_2) \ \rangle \
= \ \label{decoupling26} \\
&& {1 \over 48 \pi^2}  (X_R^i-X_L^i)^{(H)}  [ 2 (X_L^j X_L^k +
X_R^j X_R^k) +  X_L^j X_R^k + X_R^j X_L^k ]^{(H)} \epsilon^{\mu
\nu \alpha \beta} (k_1)_{\alpha} (k_2)_{\beta} \ . \nonumber \ea
When subtracted in order to define the real anomaly, the result in
(\ref{decoupling24}) is multiplied by a factor of two and can
correctly be cancelled by the anomalies of the massless (light)
fermionic spectrum, by using the initial anomaly cancellation
conditions (\ref{decoupling02}). Although $i M_H
(X_R^i-X_L^i)^{(H)} \ {\bar \psi}^{(H)} \gamma_5 \psi^{(H)}$ is
not an operator in the effective theory after integrating out the
heavy fermions, its effects can be accounted for by doubling the
axionic couplings (\ref{decoupling22}). In doing so, the anomalies
are cancelled up to local GCS terms. As we already discussed, the
coefficient of the GCS terms can always be changed by scheme
redefinition. The simplest scheme is the one in which the anomaly
is democratically distributed among the anomalous currents in the
light fermionic loops. In this scheme, the GCS terms and axionic
couplings are given by
\bea
&& E_{ij,k} \ = \ {1 \over 2} \sum_H (X_L^i X_R^j- X_R^i X_L^j)^{(H)} (X_R^k+ X_L^k)^{(H)} \ , \nonumber \\
&& C^I_{ij} \ = \  {1 \over 2 v_I} \ \sum_{H_I} [ 2 (X_L^i X_L^j + X_R^i
X_R^j) +  X_L^i X_R^j + X_R^i X_L^j ]^{(H_I)} \ . \label{decoupling27}
\eea
It is important to emphasize that, while the GCS terms are
scheme dependent due to (\ref{decoupling017}), the axionic couplings
(\ref{decoupling27}) are UV finite and therefore scheme independent.

Up to now, we have discussed in this section only massive gauge
fields, which look superficially anomalous at low energy due to
our ignorance about the high-energy anomalous set of heavy
fermions $\psi^{(H)}$. The formalism easily incorporates massless
and non-anomalous gauge bosons $A_m$, which are defined by the
necessary (but not sufficient) condition that the Higgs fields
$\phi_I$ be neutral $X_I^m = 0$, which implies in our
renormalizable examples that the heavy fermions are non-chiral
$X_L^m = X_R^m$. As a result one has $E_{mn,p}= 0$ in agreement
with our previous findings.

 In conclusion, the decoupling of heavy chiral fermions
by large Yukawa couplings does generate a generalized
Green-Schwarz mechanism at low energy, with axionic couplings
canceling anomalies of the light fermionic spectrum. It  also leads
to generalized Chern-Simons terms which play an important role in
anomaly cancellation, in analogy to the string orientifold models
we analyzed in the previous sections.

A very important and interesting question is the comparison of the
results of this section with the string theory results of the
previous sections and try to find possible differences. If possible,
this would be a remarkable way to distinguish between low-energy
predictions of string theory versus 4d field theory models. We were
not able however to find such a difference. Our present conclusion
is therefore that any experimental signature like anomalous
three-boson couplings at low energy is a strong hint towards, either
an underlying string theory with generalised anomaly cancellation
mechanism, or a standard renormalizable field theory with very heavy
chiral fermions, which generate a similar anomaly cancellation
pattern.

\section{Three gauge boson amplitudes}

\bea
\nn\\
\raisebox{-5.5ex}[0cm][0cm]{\unitlength=.65mm
\begin{fmffile}{ThreeBosonBlob_AAA11}
\begin{fmfgraph*}(30,30)
\fmfpen{thick} \fmfleft{o1} \fmfright{i1,i2}
\fmf{boson}{i1,v1,i2}\fmf{boson}{o1,v1} \fmffreeze
\fmfv{decor.shape=circle,decor.filled=shaded, decor.size=.45w}{v1}
\fmflabel{$A^\m_i(k_3)$}{o1} \fmflabel{$A^\n_j(k_1)$}{i1}
\fmflabel{$A^\r_k(k_2)$}{i2}
\end{fmfgraph*}
\end{fmffile}}
~~~~~~&=& ~~~~~~~
\raisebox{-4ex}[0cm][0cm]{\unitlength=0.4mm
\begin{fmffile}{AiAjAk_AAA11}
\begin{fmfgraph*}(60,40)
\fmfpen{thick} \fmfleft{ii0,ii1,ii2} \fmfstraight \fmffreeze
\fmftop{ii2,t1,t2,t3,oo2} \fmfbottom{ii0,b1,b2,b3,oo1}
\fmf{phantom}{ii2,t1,t2} \fmf{phantom}{ii0,b1,b2}
\fmf{phantom}{t1,v1,b1} \fmf{phantom}{t2,b2} \fmf{phantom}{t3,b3}
\fmffreeze
\fmf{photon}{ii1,v1}
\fmf{fermion,label=$p+k_2$,l.side=right}{t3,v1}
\fmf{fermion,label=$p$,l.side=right}{b3,t3}
\fmf{fermion,label=$p-k_1$,l.side=right}{v1,b3}
\fmf{photon}{b3,oo1} \fmf{photon}{t3,oo2}
\fmflabel{$A^\m_i$}{ii1} \fmflabel{$A^\n_j$}{oo1}
\fmflabel{$A^\r_k$}{oo2}
\end{fmfgraph*}
\end{fmffile}}
~~~~+~~~~~~
\raisebox{-3.7ex}[0cm][0cm]{\unitlength=.5mm
\begin{fmffile}{GCS_AAA17}
\begin{fmfgraph*}(30,30)
\fmfpen{thick} \fmfleft{o1} \fmfright{i1,i2}
\fmf{boson}{i1,v1,i2}\fmf{boson}{o1,v1} \fmflabel{$A^\m_i$}{o1}
\fmflabel{$A^\n_j$}{i1} \fmflabel{$A^\r_k$}{i2}
\end{fmfgraph*}
\end{fmffile}}
\nn\\ \nn\\ \nn\\ \nn\\ \nn\\
&&+~~\hskip -1.2cm
\raisebox{-3.5ex}[0cm][0cm]{ \unitlength=0.7mm
\begin{fmffile}{axions_AAA11}
\begin{fmfgraph*}(60,20)
\fmfpen{thick} \fmfleft{i0} \fmfright{o1,o2}
\fmf{phantom}{i0,ii1}\fmf{phantom}{ii1,i1} \fmf{photon}{i1,v0}
\fmf{plain,label=$\a_I$,l.side=left}{v0,v1} \fmf{photon}{v1,o2}
\fmf{photon}{v1,o1} \fmffreeze
\fmflabel{$A^\m_i$}{i1}\fmflabel{$A^\n_j$}{o1}
\fmflabel{$A^\r_k$}{o2}
\end{fmfgraph*}
\end{fmffile}}
~~+~~~~~~~~~\cdots \label{FigureDiagrams}\\ \nn\\ \nn\eea

There are three diagrams  (\ref{FigureDiagrams}) to evaluate : the
anomalous triangle diagrams, the tree-level axionic exchange ones
and the ones coming from the contact GCS terms.
 In the following we define: $t^{ijk}=\sum_f[Q^i_f Q^j_f Q^k_f]$.
The triangle amplitude in (\ref{FigureDiagrams}), in momentum
space, is given by
\bea
\G^{ijk}_{\m\n\r}|_{1-loop}=i^3t^{ijk}\int{d^4 p\over (2\pi)^4}
{Tr[\g_\m(\sla{p}+\sla{k_2})\g_\r \sla{p}\g_\n
(\sla{p}-\sla{k_1})\g_5] \over (p+k_2)^2(p-k_1)^2 p^2}
\eea
and can be decomposed according to
\bea
\G^{ijk}_{\m\n\r}|_{1-loop}&=& t^{ijk} [A_1(k_1,k_2) \e_{\m\n\r\s}
k_2^\s + A_2(k_1,k_2) \e_{\m\n\r\s} k_1^\s
+B_1(k_1,k_2) k_{2\n} \e_{\m\r\s\t} k_2^\s k_1^\t\nn\\
&&~~+B_2(k_1,k_2) k_{1\n} \e_{\m\r\s\t} k_2^\s k_1^\t + B_3(k_1,k_2)
k_{2\r} \e_{\m\n\s\t}k_2^\s k_1^\t + B_4(k_1,k_2) k_{1\r}
\e_{\m\n\s\t}k_2^\s k_1^\t ] \nn\\
\label{1loop}\eea
where $A$'s and $B$'s functions of $k_1,k_2$.

The coefficients $A_i,B_i$ are computed in Appendix D. The three
different contributions are given by
\bea \G^{ijk}_{\m\n\r}=\G^{ijk}_{\m\n\r}|_{1-loop}
+\G^{ijk}_{\m\n\r}|_{axion} +\G^{ijk}_{\m\n\r}|_{CS} \ , \eea
where
\bea
\G^{ijk}_{\m\n\r}|_{1-loop}&=& t^{ijk} [ A_1 \e_{\m\n\r\s} k_2^\s +
A_2 \e_{\m\n\r\s} k_1^\s
+B_1 k_{2\n} \e_{\m\r\s\t} k_2^\s k_1^\t+B_2 k_{1\n} \e_{\m\r\s\t} k_2^\s k_1^\t\nn\\
&&~~+B_3 k_{2\r} \e_{\m\n\s\t}k_2^\s k_1^\t + B_4 k_{1\r}
\e_{\m\n\s\t} k_2^\s k_1^\t] \ , \\
\G^{ijk}_{\m\n\r}|_{axion}&=& -M^i_I C^{jk}_I \left({k_{3\m} \over
k_3^2}\right) \e_{\n\r\s\t} k_2^\s k_1^\t
-M^j_I C^{ki}_I \left({k_{1\n} \over k_1^2}\right) \e_{\r\m\t\s}
k_2^\s k_3^\t
-M^k_I C^{ij}_I \left({k_{2\r} \over k_2^2}\right) \e_{\m\n\t\s}
k_3^\s k_1^\t \nn\\
&=& -M^i_I C^{jk}_I {-(k_{1\m}+k_{2\m}) \over
(k_1+k_2)^2}\e_{\n\r\s\t} k_2^\s k_1^\t
+M^j_I C^{ki}_I {k_{1\n} \over k_1^2} \e_{\m\r\s\t} k_2^\s k_1^\t
-M^k_I C^{ij}_I {k_{2\r} \over k_2^2} \e_{\m\n\s\t} k_2^\s k_1^\t
\nn\\
\\
\G^{ijk}_{\m\n\r}|_{CS}&=& -E^{ij,k} \e_{\m\n\r\s} k_2^\s
-E^{jk,i} \e_{\n\r\m\s} k_3^\s -E^{ki,j} \e_{\r\m\n\s} k_1^\s\nn\\
&=& -(E^{ij,k}-E^{jk,i}) \e_{\m\n\r\s} k_2^\s
-(E^{ki,j}-E^{jk,i}) \e_{\m\n\r\s} k_1^\s \ .
\eea

As shown in Appendix D, by using the anomaly cancellation conditions
(\ref{vanishing}), we can eliminate the scheme-dependent
coefficients $A_i$ in terms of the finite and unambiguous
coefficients $B_i$. The final result is
\ba \G^{ijk}_{\m\n\r} \ &=& \ \ \ \left[ - t_{ijk} ({C_A \over 3} +
k_1 k_2 B_1 + k_1^2 B_2) -
 E^{ij,k}+ E^{jk,i} \right] \e_{\m\n\r\s} k_2^{\s} \nonumber \\
&& + \left[  t_{ijk} ({C_A \over 3} + k_1 k_2 B_1 - k_2^2 B_3) -
 E^{ki,j}+ E^{jk,i} \right] \ \e_{\m\n\r\s} \ k_1^{\s} \nonumber \\
&& + \left[ t_{ijk} ({C_A \over 3} {k_{1\n} \over k_1^2} + k_{1\n}
B_2 +
k_{2\n} B_1) + E^{ij,k}- E^{jk,i} {k_{1\n} \over k_1^2} \right] \ \e_{\m\n\s\t} \ k_2^{\s} k_1^{\t} \nonumber \\
&& + \left[ t_{ijk} (-{C_A \over 3} {k_{2\r} \over k_2^2} - k_{1\r}
B_1 + k_{2\r} B_3) + E^{ki,j}- E^{jk,i} {k_{2\r} \over k_2^2}
\right] \ \e_{\m\n\s\t} \ k_2^{\s} k_1^{\t}  \label{total} \ . \ea

The upshot of our analysis is that in the case of a generalized
anomaly cancellation mechanism, there are anomalous three-gauge
boson couplings at low energy (\ref{total}). These couplings
involve at least one massive, anomalous gauge fields, which we
call generically $Z'$ in what follows. These new couplings could
be tested at LHC if the masses of the anomalous , $Z'$ gauge
bosons, are small enough, in the TeV range. This is possible in
orientifold models, but not only \cite{gcs6}, especially in the
case of a low fundamental string scale. The best signature of
these anomalous couplings are the $Z' \rightarrow Z \ \gamma$
decays, which to our knowledge were never considered in
phenomenological Z' models. A more detailed analysis is clearly
needed in to study the experimental consequences of these decays
in future collider experiments and particularly at LHC.

There is an interesting way to analyze the effect of GCS terms in
the CP-odd part of the three gauge-boson amplitude, in terms of its
analytic structure. Following Coleman and Grossman \cite{cg}, for
the simple kinematical configuration

\be k_1^2 \ = k_2^2 \ = \ k_3^2 \
\equiv \ Q^2 \ , \ee
the one-loop triangle contribution to the
3-boson amplitude is
\be \G^{ijk}_{\m\n\r}|_{1-loop} = - {1 \over 3
Q^2} \ t^{ijk} C_A \ [ \ \e_{\n\r\s \t} (k_1^{\m} + k_2^{\m}) +
\e_{\m\r\s\t} k_1^{\n} - \e_{\m\n\s\t} k_2^{\r} \ ] \ k_2^{\s}
k_1^\t \ . \ee
By adding the triangle diagram, the axionic exchange
which are both {\it non-local} and the {\it local} GCS contributions
and after using the gauge invariance conditions, we find the total
result
\ba
\G^{ijk}_{\m\n\r}|_{total} &=& \G^{ijk}_{\m\n\r}|_{CP=even} \nonumber \\
&& +{1 \over Q^2} \left[  \e^{\n\r\s\t} (E_{ki,j}-E_{ij,k}) (k_1^{\m}
+ k_2^{\m}) +
\e^{\m\r\s\t} (E_{ij,k}-E_{jk,i}) k_1^{\n} \right. \nonumber \\
&& \left. +  \e^{\m\n\s\t} (E_{ki,j}-E_{jk,i}) k_2^{\r}  \right] k_2^{\s}  k_1^{\t} \nonumber \\
&& -  E_{ij,k} \e^{\m\n\r\s}  k_2^{\s} + E_{jk,i} \e^{\n\r\m\s}
(k_1^{\s} + k_2^{\s}) -  E_{ki,j} \e^{\r\m\n\s}  k_1^{\s} \ . \ea
Notice that the pole in $1/Q^2$ is completely determined by the GCS
terms and does not exist if they are absent.

\vskip 2cm
\begin{flushleft}
{\large \bf Acknowledgments}
\end{flushleft}
\vskip 2cm

\noindent It is a pleasure to thank Luis Alvarez-Gaume, Sergio
Ferrara, Boris Kors, Costas Kounnas, Augusto Sagnotti, Timo
Weigand and Arkady Vainshtein for useful discussions. M. B. and E.
D. would like to thank the Galileo Galilei Institute for
Theoretical Physics for hospitality and INFN for partial support
during completion of this work.
This work was supported in part by the CNRS PICS no. 2530 and
3059, INTAS grant 03-516346, MIUR-COFIN 2003-023852, NATO
PST.CLG.978785, the RTN grants MRTN-CT-2004-503369, EU
MRTN-CT-2004-512194, MRTN-CT-2004-005104 and by a European Union
Excellence Grant, MEXT-CT-2003-509661. P. A. was supported by the
research program ``Pythagoras II'' (grant 70-03-7992) of the Greek
Ministry of National Education, partially funded by the European
Union.

 \newpage
\appendix

 \renewcommand{\theequation}{\thesection.\arabic{equation}}
\addcontentsline{toc}{section}{Appendices}
\section*{APPENDIX}

\section{Triangle anomalies and regularization dependence\label{a}}

In this appendix we will present some known facts about triangle
graphs and scheme dependence. They are useful in our general
analysis of the effective action. We use the conventions of
Weinberg's textbook, \cite{weinberg} to which we refer the reader
for all details that we omit here.

We use a basis for the fermions so that they are all left-handed.
We will package them into a single spinor $\psi$.
The associated charge operator for the gauge field $A_{\m}^i$ is
 denoted by ${\cal Q}_i$.
We define the various U(1) currents as
\be
J^{\mu}_i=-i\bar\psi {\cal Q}_i\gamma^{\mu}\psi
\label{a1}\ee
The three-current correlator we will study is
\be
\Gamma^{\mu\nu\rho}_{ijk}(x,y,z)=\langle J_i^{\mu}(x)J_i^{\nu}(y)J_i^{\rho}(z)\rangle
\label{a2}\ee
The leading contribution, at one loop emerges from fermions going around the loop.
The total contribution is obtained by summing over all relevant fermion fields.

There are two diagrams for the correlator that can be evaluated to yield
\bea
\Gamma^{\mu\nu\rho}_{ijk}(x,y,z)&=&-i{\rm Tr}\left[S(x-y)Q_j\gamma^{\n}
P_{L}S(y-z)Q_k \gamma^{\rho}P_LS(z-x)Q_i\gamma^{\m}P_L\right]
\nn\\
&&-i{\rm Tr}\left[S(x-z)Q_k\gamma^{\r}P_{L}S(z-y)Q_j \gamma^{\n}P_LS(y-x)Q_i\gamma^{\m}P_L\right]
\label{a3}\eea
with
\be
P_{L}={1+\gamma^5\over 2}\sp S(x)=-\int {d^4 p\over (2\pi)^4}{\sla{p}\over p^2}e^{ip\cdot x}
\label{a4}\ee
Substituting we obtain
\bea
\Gamma^{\mu\nu\rho}_{ijk}(x,y,z)
&=&{it_{ijk}}\int {d^4 k_1\over (2\pi)^4}
 {d^4 k_2\over (2\pi)^4}e^{-i(k_1+k_2)\cdot x+ik_1\cdot y+ik_2\cdot z}\int {d^4 p\over (2\pi)^4}
 \times\nn\\
&& \bigg\{ {\rm Tr}\left[{\sla{p}-\sla{k_1}+\sla{a}\over (p-k_1+a)^2}\gamma^{\n}
{\sla{p}+\sla{a}\over (p+a)^2}\gamma^{\r}
{\sla{p}+\sla{k_2}+\sla{a}\over (p+k_2+a)^2}\gamma^{\m}P_L\right]+\nn\\
&&+{\rm Tr}\left[{\sla{p}-\sla{k_2}+\sla{b}\over (p-k_2+b)^2}\gamma^{\r}
{\sla{p}+\sla{b}\over (p+b)^2}\gamma^{\n}
{\sla{p}+\sla{k_1}+\sla{b}\over (p+k_1+b)^2}\gamma^{\m}P_L\right]\bigg\}
\label{a5}\eea
with $t_{ijk}={\rm Tr}[{\cal Q}_i{\cal Q}_j{\cal Q}_k]$
We have shifted the integrated momentum in the two diagrams using two vectors $a_{\m}$ and $b_{\m}$.
This reflects the standard ambiguity of the triangle graph and translates into the
 definition of the associated current opeartors.
Demanding that there is no anomaly in the vector currents
forces $b=-a$, choice that we keep from now on.
The vector $a$ is parameterizing the leftover scheme dependence of the triangle graph in question.

We may now obtain the following divergence formulae,
\be
\partial_{\mu}\Gamma^{\mu\nu\rho}_{ijk}(x,y,z)=-{t_{ijk}\over 8\pi^2}\int {d^4 k_1\over (2\pi)^4}
 {d^4 k_2\over (2\pi)^4}e^{-i(k_1+k_2)\cdot x+ik_1\cdot y+ik_2\cdot z}~\epsilon^{\nu\rho\s\t}~a_{\s}(k_1+k_2)_{\t}
\label{a6}\ee
\be
\partial_{\n}\Gamma^{\mu\nu\rho}_{ijk}(x,y,z)=-{t_{ijk} \over 8\pi^2}\int {d^4 k_1\over (2\pi)^4}
 {d^4 k_2\over (2\pi)^4}e^{-i(k_1+k_2)\cdot x+ik_1\cdot y+ik_2\cdot z}~\epsilon^{\m\rho\s\t}~(a+k_2)_{\s}(k_1)_{\t}
\label{a7}\ee
\be
\partial_{\r}\Gamma^{\mu\nu\rho}_{ijk}(x,y,z)=-{t_{ijk} \over 8\pi^2}\int {d^4 k_1\over (2\pi)^4}
 {d^4 k_2\over (2\pi)^4}e^{-i(k_1+k_2)\cdot x+ik_1\cdot y+ik_2\cdot z}~\epsilon^{\m\n\s\t}~(k_1-a)_{\s}(k_2)_{\t}
\label{a8} \ee
A generic choice of scheme (i.e. $a_{\m}$) indicates that the divergence structure is asymmetric among the
three vertices of the triangle graph.
There is a single choice that is fully symmetric, namely
\be
a={1\over 3}(k_1-k_2)
\label{a9}\ee

We now proceed to construct the effective action for the gauge fields after integrating out the fermions.
To cubic order we obtain
\be
S_{ijk}={1\over 3!}\int d^4x ~d^4y ~d^4 z~ \Gamma^{\mu\nu\rho}_{ijk}
(x,y,z)~A_{\mu}^i(x)~A^j_{\n}(y)~A^k_{\r}(z)
\label{a10}\ee
where no summation is assumed on the $i,j,k$ labels.

Upon gauge transformations $A_{\mu}^i\to A_{\m}^i+\partial_{\mu}\varepsilon^i$ we obtain
\be
\delta S_{ijk}=-{1\over 3!}\int~\left[\varepsilon^i \partial_{\mu}\Gamma^{\mu\nu\rho}_{ijk}
A^j_{\n}(y)~A^k_{\r}(z)+\varepsilon^j \partial_{\nu}\Gamma^{\mu\nu\rho}_{ijk}
A^i_{\m}(x)~A^k_{\r}(z)+\varepsilon^k \partial_{\r}\Gamma^{\mu\nu\rho}_{ijk}
A^i_{\m}(x)~A^j_{\n}(y)\right]
\label{a11}\ee

If $a$ is a constant independent of momenta then it does not contribute to the gauge variations.
We therefore parameterize the scheme dependence as
\be
a= A k_1+ B k_2
\label{a12}\ee
The real numbers $A,B$ can be different for different ijk combinations.
\ba
\delta S_{ijk} \ &=& \ -{t_{ijk}\over 3!(32\pi^2)} \int d^4x~ \Big\{  {(A-B)_{ijk}}\varepsilon^i ~
\epsilon^{\m\n\r\s}~F^j_{\m\n}F^k_{\r\s}
+{(B_{ijk}+1)}\varepsilon^j ~\epsilon^{\m\n\r\s}~F^i_{\m\n}F^k_{\r\s}   \nonumber \\
&& \ \ \ \ \ \ \ \ \ \ \ \ \ \ \ \ \ \ \ \ \ \ \ \ \
- {(A_{ijk}-1)}\varepsilon^k ~\epsilon^{\m\n\r\s}~F^i_{\m\n}F^j_{\r\s} \Big\} \ . \label{generals}
\label{a13}\ea

We will now fix the symmetric scheme $A_{ijk}=-B_{ijk}=1/3$ in which we obtain the gauge variation
\ba
&& \delta S_{ijk} \ = \ - {t_{ijk}\over 3!(12\pi^2)}\int d^4x~
\Big\{ \varepsilon^i ~
~F^j\wedge F^k
 +\varepsilon^j ~~F^i\wedge F^k+ \varepsilon^k ~
 ~F^i\wedge F^j \Big\}
\label{a14}\ea
where we used
\be
 F^i\wedge F^j={1\over 4}\epsilon^{\m\n\r\s}~F^i_{\m\n}F^j_{\r\s}
\label{a15}\ee

We now sum over the U(1)'s to obtain the full cubic effective action in the symmetric scheme.
Its gauge variation is
\be
\delta S_3=\sum_{i,j,k}\delta S_{ijk}=-{t_{ijk}\over 24\pi^2}\int d^4x~ \ve^i F^j\wedge F^k
\label{a015}\ee
where we have reinstated our summation convention.

We may now study the effect of changing the scheme of the triangle graphs.
This is obtained by setting
\be
A_{ijk}={1\over 3}+\tilde A_{ijk}\sp B_{ijk}=-{1\over 3}+\tilde B_{ijk}
\label{a16}\ee
The gauge variation now becomes
\bea
 \delta S_3 &=&- {t_{ijk}\over 24\pi^2}\int d^4x~\Big\{ \varepsilon^i ~
~F^j\wedge F^k\Big\}\label{a17}\\
&&-{t_{ijk}\over 3!(8\pi^2)} \int d^4x \Big\{\tilde A_{ijk}(\varepsilon^i F^j\wedge F^k-\varepsilon^k F^i\wedge F^j)
-\tilde B_{ijk}(\varepsilon^i F^j\wedge F^k-\varepsilon^j F^i\wedge F^k)\Big\}
\nn\eea

The extra terms have the same transformation properties as
\be
S_{\rm counter}= -{t_{ijk}\over 3!(16\pi^2)} \int d^4x~\epsilon^{\m\n\r\s}
\left[\tilde A_{ijk} ~A^i_{\m}A^k_{\n}F^j_{\r\s}
-\tilde B_{ijk} ~A^i_{\m}A^j_{\n}F^k_{\r\s}\right]
\label{a18}\ee
Therefore, in this new scheme, the new effective action is obtained from the old one by adding the GCS terms
in (\ref{a18}).

A more direct way to see this is to compute the variation of the effective action between two
different regularisation schemes specified by the shift vectors
$a_1^{ijk} $ and $a_2^{ijk}$, where $a^{ijk} = A_{ijk} k_1 + B_{ijk} k_2 $ :
\be
\Delta \Gamma^{\mu \nu \rho}_{ijk} (x,y,z) = \Gamma^{\mu \nu \rho}_{ijk}|_{a_1} - \Gamma^{\mu \nu \rho}_{ijk}|_{a_2}  \  .  \label{tr3}
\ee
By Taylor expanding
\ba
\Delta \Gamma ^{\mu \nu \rho} (p,k_i,a) &=& (a_2-a_1)^{\sigma} {\partial \over \partial a_1^{\sigma}} \Gamma^{\mu \nu \rho} (p,k_i,a_1) \nonumber \\
&&+{1 \over 2} (a_2-a_1)^{\sigma_1} (a_2-a_1)^{\sigma_2} {\partial^2 \over \partial a_1^{\sigma_1} \partial a_2^{\sigma_2}}  \Gamma^{\mu \nu \rho} (k_i,a_1) + \cdots  \label{tr4}
\
\ea
and noticing that ${\partial  \Gamma ^{\mu \nu \rho} (p,k_i,a)  /  \partial a^{\sigma}} {\partial  \Gamma ^{\mu \nu \rho} (p,k_i,a)  /  \partial p^{\sigma}}$, we can cast the scheme
difference into the form
\ba
\Delta \Gamma^{\mu \nu \rho}_{ijk} (x,y,z) &=& {i \over (2 \pi)^{12}} \int d^4 k_1 d^4 k_2
\ e^{-i (k_1+k_2)x + i k_1 y + i k_2 z}  (a_2-a_1)^{\sigma}   \times \nonumber \\
&& \int d^4 p \
t_{ijk}  {\partial \over \partial p^{\sigma}}  \left[  \Gamma^{\nu \rho \mu}_{ijk} (p,k_1,k_2,a_1)  -  \Gamma^{\rho \nu \mu}_{ijk} (p,k_2,k_1,-a_1) \right]  \  + \cdots \ ,~~~~~~~
\label{tr5}
\ea
where $\cdots$ are contributions at least quadratic in the shift vectors $a$ containing at least
second derivatives with respect to the loop momentum $p$. Since all contributions come
from the boundary of the loop momentum space, we will see in a moment that only
the first contribution gives a non-vanishing contribution.
Like in the case of the triangle gauge anomalies, the quantity
$ \Delta \Gamma^{\mu \nu \rho}_{ijk}$ is given by a surface contribution.
A simple counting of the leading momentum dependence for $p \rightarrow \infty$
shows that only the leading contribution
\be
\Gamma^{\nu \rho \mu}_{ijk} (p,k_1,k_2,a)  \rightarrow   - {2 \over p^6}
\left[  p^2 (p^{\mu} \eta^{\nu \rho} + p^{\nu} \eta^{\mu \rho} + p^{\rho} \eta^{\mu \nu }) -
4 p^{\mu}  p^{\nu}  p^{\rho}  + i p^2 \epsilon^{\nu \rho \mu \sigma}  p_{\sigma}  \right]  \
\label{tr6}
\ee
is giving a non-vanishing result and only the last term in (\ref{tr6}) contributes to (\ref{tr5}).
By explicitly computing now the surface integral
\be
\int d^4 p \ \partial_{\sigma} {p_{\epsilon} \over p^4} = - {1 \over 8}  \eta_{\sigma \epsilon}
\int d^4 p  \ \partial^2 {1 \over p^2} = - {\pi^2 \over 4}    \eta_{\sigma \epsilon}   \ ,  \label{tr7}
\ee
we finally get the difference of the effective action in two different regularisation schemes
to be equal to
\ba
\Delta S_3^{\rm an}  &=&    {1\over 3!}\int d^4x ~d^4y ~d^4 z~  \Delta \Gamma^{\mu\nu\rho}_{ijk}
(x,y,z)~A_{\mu}^i(x)~A^j_{\n}(y)~A^k_{\r}(z) \nonumber \\
&=&  {1 \over 32 \pi^2} t_{ijk}  \  \left( \Delta A_{ikj} - \Delta B_{ijk} \right)
\int  \ A^i \wedge A^j \wedge F^k  \  . \label{tr8}
\ea

We will do hear a counting of the relevant parameters.
we start with all possible independent GCS terms ${\cal S}_{ijk}$
constructed out of N abelian gauge bosons.
The relations are, antisymmetry in the first two indices as well as cyclic symmetry
\be
{\cal S}_{ijk}+{\cal S}_{jik}=0\sp {\cal S}_{ijk}+{\cal S}_{jki}+{\cal S}_{kij}=0
\label{a19}\ee
We have to distinguish the following cases:

iii) Then the GCS term is trivial

ijj) $i\not= j$. There are two possible GCS terms per pair of distinct gauge bosons, namely
${\cal S}_{ijj}$ and ${\cal S}_{jii}$. This gives a total of $N(N-1)$ independent terms.

ijk) with $i\not= j\not= k\not= i$. Here out of the three possible terms only two are independent.
The third is related to the other two by the cyclicity property in (\ref{a19}).
We therefore obtain here ${N(N-1)(N-2)\over 3}$ independent terms.

Therefore the total number is ${N(N^2-1)\over 3}$ corresponding to the Young tableau
$\Yboxdim5pt\yng(2,1)$.

It would naively seem that this number is smaller than the number of possible schemes,
specified by the coefficients  $\tilde A_{ijk}$, $\tilde B_{ijk}$, namely , $2N^3$.
We will now show that the number of relevant scheme parameters is exactly equal to the number of
independent GCS terms.

iii) For this case, we must choose $\tilde A_{iii}=\tilde B_{iii}=0$ to respect the full Bose symmetry
of the triangle graph.

iij) $i\not= j$. In this case, (\ref{a17}) indicates that  the scheme depends only on $\tilde A-\tilde B$,
and there are N(N-1) such coefficients.

ijk)  with $i\not= j\not= k\not= i$. For this case we have $2 \times {N(N-1)(N-2)\over 3!}$
such coefficients.

Therefore the scheme dependence of triangle graphs is in one to one correspondence with all possible GCS terms.

We now split the U(1)s into two groups, as was done in section \ref{general}.
We also add the non-abelian mixed graphs. Using (\ref{a14})  we obtain
the gauge variation of the effective action due to the triangle graphs in the symmetric scheme,
\bea
\delta S_{\rm triangle}=-{1\over 24\pi^2}\int &\Big\{&
t_{abc}\ve^a ~F^b\wedge F^c+t_{mnr}\ve^m ~F^n\wedge F^r
\label{a20}\\
&&+t_{mab}(2\ve^a ~F^b\wedge F^m+\ve^m ~F^a\wedge F^b)\nn\\
&&+
t_{amn}(2\ve^m ~F^a\wedge F^n+\ve^a ~F^m\wedge F^n)\nn\\
&&
+{T^{a}}_{\a}(2{\rm Tr}[\ve~\tilde G^{\a}]\wedge F^a+\ve^a ~{\rm Tr}[G^{\a}\wedge G^{\a}])
\nn\\
&&+{T^m}_{\a}(2{\rm Tr}[\ve~\tilde G^{\a}]\wedge F^m+\ve^m ~{\rm Tr}[G^{\a}\wedge G^{\a}])
\Big\}
\nn\eea
The tensor  $T$ is given by the cubic traces of the U(1) and non-abelian generators,
\be
{T^{a}}_{\a}={\rm Tr}[{\cal Q}_a(TT)^{\a}]\sp {T^m}_{\a}={\rm Tr}[{\cal Q}_m(TT)^{\a}]
\label{a21}\ee
with $(TT)^{\a}$ the quadratic Casimir of the $\a$-th non-abelian factor.

\section{Basis changes\label{b}}

In this appendix we relate effective couplings in the D-brane basis, as calculated in string theory
and the diagonal basis, where the gauge-boson mass-matrix is diagonal.
 This basis was introduced in detail in section
\ref{general}, (\ref{4})-(\ref{11}).
We obtain the following equations relating the PQ  couplings
\be
{C^M}_{ab}=W^I_{M}\eta^a_{i}\eta^b_{j}{C^I}_{ij}
\sp
{C^M}_{mn}=W^I_{M}\eta^m_{i}\eta^n_{j}{C^I}_{ij}
\sp
{C^M}_{am}=2W^I_{M}~\eta^a_{i}\eta^m_{j}~{C^I}_{ij}
\label{b1}\ee
\be
{C^a}_{bc}=W^I_{a}M_a\eta^b_{i}\eta^c_{j}{C^I}_{ij}
\sp
{C^a}_{mn}=W^I_{a}M_a\eta^m_{i}\eta^n_{j}{C^I}_{ij}
\sp
{C^a}_{bm}=2W^I_{a}M_a~\eta^b_{i}\eta^m_{j}~{C^I}_{ij}
\label{b2}\ee
\be
{D^M}_{\a}=W^I_M~ {C^I}_{\a}\sp {D^a}_{\a}=W^I_a~M_a~ {C^I}_{\a}
\label{b3}\ee
\be
{C^I}_{ij}={M^I_k\eta^a_k\over M_a^2}\left[{C^a}_{bc}\eta^b_i\eta^c_j+
{1\over 2}{C^a}_{bm}(\eta^m_i\eta^b_j+\eta^m_j\eta^b_i)+{C^a}_{mn}\eta^m_i\eta^n_j\right]+
\label{b4}\ee
$$
+W^I_M\left[{C^M}_{bc}\eta^b_i\eta^c_j+
{1\over 2}{C^M}_{mb}(\eta^m_i\eta^b_j+\eta^m_j\eta^b_i)+{C^M}_{mn}\eta^m_i\eta^n_j\right]
$$

\be
M^I_i{C^I}_{jk}=\eta^a_i\left[{C^a}_{bc}\eta^b_j\eta^c_k+
{1\over 2}{C^a}_{bm}(\eta^m_j\eta^b_k+\eta^m_k\eta^b_j)+{C^a}_{mn}\eta^m_j\eta^n_k\right]
\label{b5}\ee
where we used
\be
M^I_i W^I_{M}=0
\label{b6}\ee
the GCS couplings
\be
E_{abc}=\eta^a_{i}\eta^b_{j}\eta^c_{k} ~E_{ijk}
\sp E_{mnr}=\eta^m_{i}\eta^n_{j}\eta^r_{k} ~E_{ijk}
\label{b7}\ee
\be
E_{man}=2(\eta^m_i\eta^a_j\eta^n_{k}+\eta^m_i\eta^n_j\eta^a_{k})E_{ijk}
\label{b8}\ee
\be
E_{mab}=2(\eta^m_i\eta^a_j\eta^b_{k}-\eta^a_i\eta^b_j\eta^m_k)E_{ijk}
\label{b9}\ee
\be
{Z^a}_{\a}=\eta^a_i~{Z^i}_{\a}\sp {Z^m}_{\a}=\eta^m_i~{Z^i}_{\a}
\label{b10}\ee
\be
E_{ijk}=\eta^m_i\eta^n_j\eta^r_k~E_{mnr}+
{1\over 2}(\eta^m_i\eta^a_j-\eta^m_j\eta^a_i)\eta^n_k~E_{man}+
{1\over 2}(\eta^m_i\eta^a_j-\eta^m_j\eta^a_i)\eta^b_k~E_{mab}+
\eta^a_i\eta^b_j\eta^c_k~E_{abc}
\label{b11}\ee
For the charges, we start from the coupling
\be
S_{minimal}=\int ~\bar\psi ~{\cal Q}_i~A^i_{\m}\gamma^{\mu}~\psi
\label{b12}\ee
and by changing basis this becomes
\be
S_{minimal}=\int ~\left[\bar\psi ~{\cal Q}_a~Q^a_{\m}\gamma^{\mu}~
\psi+\bar\psi ~{\cal Q}_m~Y^m_{\m}\gamma^{\mu}~\psi\right]
\label{b13}\ee
with
\be
{\cal Q}_a=\eta^a_{i}{\cal Q}_i\sp {\cal Q}_m\eta^m_{i}{\cal Q}_i
\label{b14}\ee
Therefore
\be
t_{abc}=\eta^a_i\eta^b_j\eta^c_k ~t_{ijk}
\sp
t_{abm}=\eta^a_i\eta^b_j\eta^m_k ~t_{ijk}
\sp
t_{amn}\eta^a_i\eta^m_j\eta^n_k ~t_{ijk}
\label{b15}\ee
It is also convenient to introduce the projections
\be
G^{ij}\equiv\eta^a_i\eta^a_{j}\sp \tilde G^{ij}\equiv\eta^m_i\eta^m_{j}\sp G^{ij}+\tilde G^{ij}=\delta^{ij}
\label{b16}\ee
$G$ projects on the subspace of massive U(1)s while $\tilde G$ to the massless one,
\be
G^{ij}G^{jk}=G^{ik}\sp \tilde G^{ij}\tilde G^{jk}=\tilde G^{ik}
\sp \tilde G^{ij} G^{jk}=G^{ij} \tilde G^{jk}=0
\label{b19}\ee
We also define
\be
\tilde M_{ij}\equiv {1\over M^2_a}\eta^a_i\eta^a_j
\label{b17}\ee
This is the inverse of $M^2_{ij}$ in the invertible (massive) subspace.
It satisfies
\be
\tilde M_{ij}\eta^a_j={1\over M_a^2}\eta^a_i\sp \tilde M_{ij}\eta^m_j=0
\label{b18}\ee

\TABLE[t]{\footnotesize
\renewcommand{\arraystretch}{1.25}
\begin{tabular}{|c|c|c|}
\hline
Twist Group   & & \\
\cline{1-1} Gauge Group & \raisebox{2.5ex}[0cm][0cm]{ (99)/(55)
matter} &
\raisebox{2.5ex}[0cm][0cm]{ (95) matter}  \\
\hline\hline $Z_6 $ & $2 (15,1,1)+ 2(1,\overline{15},1) $ &
$(6,1,1;6,1,1)+(1,\overline{6},1;1,\overline{6},1)$  \\
\cline{1-1} $U(6)_9^2\times U(4)_9\times$ & $ +2
(\overline{6},1,4)+ 2(1,6,\overline{4}) $ &
$ +(1,6,1;1,1,\overline{4})+(1,1,\overline{4};1,6,1) $\\
$U(6)^2_5\times U(4)_5$ & $ +(\overline{6},1,\overline{4}) +
(1,6,4) + (6,\overline{6},1) $ &
$+(\overline{6},1,1;1,1,4)+(1,1,4;\overline{6},1,1)$ \\
\hline\hline $Z_6' $ & $(\bar{4},1,8)+
(1,4,\bar{8})+(6,1,1)+(1,\bar{6},1)$ &
$(\bar{4},1,1;\bar{4},1,1)+(1,4,1;1,4,1)$  \\
\cline{1-1} $U(4)_9^2\times U(8)_9\times$ & $ + (4,1,8)
+(1,\bar{4},\bar{8}) +(\bar{4},4,1) + (1,1,28)$ &
$ +(1,\bar{4},1;1,1,8)+(1,1,\bar{8};1,\bar{4},1) $\\
$U(4)^2_5\times U(8)_5$ & $ + (1,1,\bar{28})
+(4,4,1)+(\bar{4},\bar{4},1)$ &
$+(4,1,1;1,1,\bar{8})+(1,1,\bar{8};4,1,1)$ \\
\hline
\end{tabular}
\caption{The transformations of the massless fermionic states in
the $Z_6$ and $Z_6'$ D=4 orientifold.}\label{Massless1}}

\section{Explicit Orientifold Examples}
Here we  evaluate explicitly the anomaly-related charge traces and
the coefficients of the generalized Chern-Simons terms for the
$Z_6$ and $Z_6'$ orientifolds. We have chosen these examples as
they contain all the non-trivial
 ingredients of a generic orientifold vacuum.

\subsection{$Z_6$ orbifold}
The orbifold rotation vector is $(v_1,v_2,v_3)=(1,1,-2)/6$. There
is an order two twist ($k=3$) and we must have one set of
D5-branes. Tadpole cancellation then implies the existence of 32
D9-branes and 32 D5-branes, that we put together at the origin of
the internal space. The Chan--Paton vectors are
\be V_9=V_5={1\over 12}(1,1,1,1,1,1,5,5,5,5,5,5,3,3,3,3) \ , \ee
giving
\be tr[\gamma_k]=0\ \ \ {\rm for}\ \ k=1,3,5\sp tr[\gamma_2]=4\sp
tr[\gamma_4]=-4\, . \ee
The gauge group has a factor of $U(6)\times U(6)\times U(4)$
coming from the D9-branes and an isomorphic factor coming from the
D5-branes. The massless spectrum is provided in table
\ref{Massless}.
The ${\cal N}=1$ sectors correspond to $k=1,2,4,5$, while $k=3$ is
an ${\cal N}=2$ sector.

\subsection*{Anomaly traces}
Here we evaluate the mixed anomaly matrixes, from the  massless
spectrum of the $Z_6$ orientifold (table \ref{Massless}). We
normalize the generators of $U(1)_i$ so that the charges are $\pm
1$ for fundamentals and $\pm 2$ for symmetric / antisymmetric
tensors. This implies that $Q_{1,2}=\lambda_{1,2}$ while
$Q_3={2\over \sqrt{6}}\lambda_3$ where $\lambda_i$ are the
generators given in (\ref{lamdas}).

We also normalize the generators of the non-abelian factors as
$Tr[T_i T_j]_{\Yboxdim5pt\yng(1)}=\delta_{ij}$ in the fundamental.
This implies for example that for $SU(N)$, the same trace gives
$Tr[T_i T_j]_{\Yboxdim5pt\yng(1,1)}=(N-2)\delta_{ij}$ for the
antisymmetric representation.

Therefore, for the mixed anomalies between abelian and non-abelian
factors we have:
\be t_{ia}\equiv Tr[Q_i(T^AT^A)_a]=\left(
\begin{array}{rrrrrr}
 6 & -3 & 2 & 3 & 0 & 2 \\
 3 & -6 & -2 & 0 & -3 & -2 \\
 -9 & 9 & 0 & -3 & 3 & 0 \\
 3 & 0 & 2 & 6 & -3 & 2 \\
 0 & -3 & -2 & 3 & -6 & -2 \\
 -3 & 3 & 0 & -9 & 9 & 0
\end{array}
\right) \ee
where the columns label the U(1)s and the rows the non-abelian
factors.
It can be directly verified that  there are three linear
combinations of the U(1)s which are free of 4d mixed non-abelian
anomalies.

For abelian mixed anomalies we must evaluate
$t_{ijk}=Tr[Q_iQ_jQ_k]$. It is enough to calculate
\be
t_{ij}=\left\{
\begin{array}{rr} t_{ijj} &~,~{\rm for}~ i\neq j\\
3t_{ijj} &~,~{\rm for}~ i= j\end{array}\right. \label{tij}\ee
because in this basis (D-brane basis), $t_{ijk}=0$ when all
$i,j,k$ are distinct (chiral fermions can carry at most two U(1)
charges).
Therefore:
\be t_{ij}=\left(
\begin{array}{rrrrrr}
216 &  -36 & 24 &  36 &   0 & 24 \\
 36 & -216 &-24 &   0 & -36 &-24 \\
-72 &   72 &  0 & -24 &  24 &  0 \\
 36 &    0 & 24 & 216 & -36 & 24 \\
  0 &  -36 &-24 &  36 &-216 &-24 \\
-24 &   24 &  0 & -72 &  72 &  0
\end{array}\right)\label{tijZ6}\ee

\subsection*{Anomalous U(1) masses}

The various contributions to the mass matrix are
\bea {1\over 2}M^2_{99,ij}&=&-{\sqrt{3}\over 48\pi^3}
\Big(tr[\gamma_1\lambda_i^9]tr[\gamma_1\lambda_j^9]+tr[\gamma_5\lambda_i^9]
tr[\gamma_5\lambda_j^9]\nn\\
&&+3(tr[\gamma_2\lambda_i^9]tr[\gamma_2\lambda_j^9]+
tr[\gamma_4\lambda_i^9]tr[\gamma_4\lambda_j^9])\Big) -{{\cal
V}_3\over
3\pi^3}tr[\gamma_3\lambda_i^9]tr[\gamma_3\lambda_j^9]~~~~
\label{M299ofZ6}\eea
and similarly for $M_{55,ij}$, while
\bea {1\over 2}M^2_{95,ij}&=&-{\sqrt{3}\over 48\pi^3}\Big(
[tr[\gamma_1\lambda_i^9]tr[\gamma_1\lambda_j^5]+
tr[\gamma_5\lambda_i^9]tr[\gamma_5\lambda_j^5]\nn\\
&&+tr[\gamma_2\lambda_i^9]tr[\gamma_2\lambda_j^5]
+tr[\gamma_4\lambda_i^9]tr[\gamma_4\lambda_j^5]\Big) -{{\cal
V}_3\over 12\pi^3}tr[\gamma_3\lambda_i^9]tr[\gamma_3\lambda_j^5]\,
. \label{M295ofZ6}\eea
This mass matrix has the following eigenvalues and eigenvectors:
\bea
\begin{tabular}{lcl}
$m_1^2=0$&,& $A_1+A_2-\tilde A_1-\tilde A_2+\sqrt{6}(A_3-\tilde
A_3)$,\\
$m_2^2={3\sqrt{3}/ 2}$&,& $A_1-A_2-\tilde A_1+\tilde A_2$,\\
$m_3^2={3\sqrt{3}}$&,& $A_1-A_2+\tilde A_1-\tilde A_2$,\\
$m_4^2={40}{\cal V}_3/3$&,& $-\sqrt{3\over 2}(A_1+A_2-\tilde
A_1-\tilde A_2)-A_3+\tilde A_3$,\\
$m^2_{\pm}={7\sqrt{3}+80{\cal V}_3\pm\sqrt{147-1040\sqrt{3}{\cal
V}_3+6400{\cal V}_3^2} \over 12}$&,& $a_{\pm}(A_1+A_2+\tilde
A_1+\tilde A_2)+A_3+\tilde A_3$
\end{tabular}\label{MassesZ6}
\eea
where $A,\tilde{A}$ denote the abelian bosons which are coming
from D9 and D5 branes respectively. Also
\be a_{\pm}={40{\cal V}_3-\sqrt{3}\pm\sqrt{147-1040\sqrt{3}{\cal
V}_3+6400{\cal V}_3^2}\over 12\sqrt{2}-40\sqrt{6}{\cal V}_3}\, .
\ee
In the limit ${\cal V}_3\to 0$ two more masses become zero ($m_4$
and $m_-$). It is straightforward to check that these  three gauge
are anomaly-free in four bosons dimensions. This behavior, was
explained in detail in \cite{akr}.

\subsection*{Generalized CS terms from the non-planar cylinder}
\begin{figure}[h]
\begin{center}
\epsfig{file=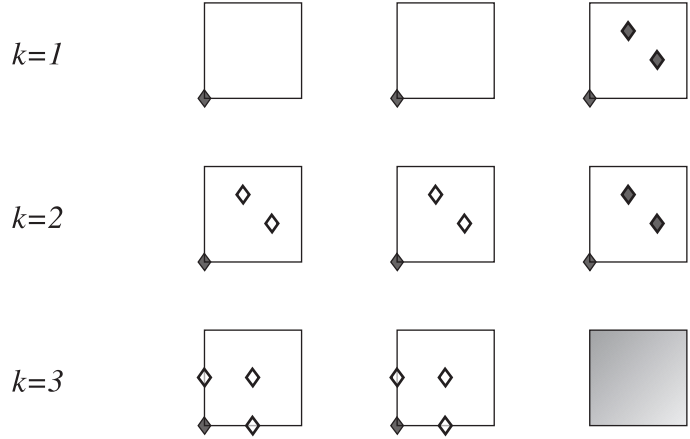,width=90mm}
\end{center}\caption{We denote by
$\blacklozenge$/$\lozenge$ the fixed points on each torus, which
are invariant/related to others by the $Z_6$ action.\label{fpZ6}}
\end{figure}

Using formulae (\ref{M299ofZ6}-\ref{M295ofZ6}), we can evaluate
the couplings of axions to one and two gauge bosons, ( $M_I$'s and
$C_I$'s respectively). Since axions (which are coming from the
twisted closed string sector) are localized at the fixed points,
it is necessary to identify the fixed points in each sector and
their properties. In  figure \ref{fpZ6} we denote the fixed points
on each torus under the $Z_6$ action.

The $k=1$ sector provides 3 points fixed under the $Z_6$ action.
The $k=2$ sector provides 27 points fixed under the $Z_3$ action.
However, the $Z_6$ action leaves invariant only 3 of them (the
ones of the $k=1$ sector) and relates doublets of the rest. In
total there are 12 doublets of points which are identified under
the $Z_6/Z_3=Z_2$ action.
The $k=3$ sector provides 16 points fixed under the $Z_2$ action.
However, the $Z_6$ action leaves invariant only 1 of them (which
is located at the origin) and relates triplets of the rest. In
total there are 5 triplets of points identified under the
$Z_6/Z_2=Z_3$ action.

Taking all this into account,  the D9 branes couple to axions as:
\bea
M_{a(9)}^{1,\circlearrowleft} &=& {i \over \sqrt{48\pi^3}}
{1\over3^{1/4}} tr[\gamma_1 \lambda_a^9]\nn\\
M_{a(9)}^{2,\circlearrowleft} &=& {i \over \sqrt{48\pi^3}}
{1\over27^{1/4}} tr[\gamma_2 \lambda_a^9]~,~~~~
M_{a(9)}^{2,\rightleftarrows} = {i \over \sqrt{48\pi^3}}
{\sqrt{2}\over27^{1/4}} tr[\gamma_2 \lambda_a^9]\nn\\
M_{a(9)}^{3,\circlearrowleft} &=& {i \sqrt{2{\cal V}_3} \over
\sqrt{24\pi^3}} {1\over16^{1/4}} tr[\gamma_3 \lambda_a^9]~,~~~~
M_{a(9)}^{3,\rightleftarrows} = {i \sqrt{2{\cal V}_3} \over
\sqrt{24\pi^3}} {\sqrt{3}\over16^{1/4}} tr[\gamma_3 \lambda_a^9]
\label{MI9sZ6}\eea
where $\circlearrowleft$ denotes fixed points of the $k$th sector
which are also fixed under the larger $Z_6$ orbifold action
(corresponding to $\blacklozenge$ on all two-tori $T^2_i$) and
$\rightleftarrows$ denotes fixed points which are related by the
larger $Z_6$ orbifold action (with $\lozenge$ in at least one tori
$T^2_i$) .

If all D5 branes are at the origin then the corresponding
couplings are:
\bea
M_{a(5)}^{1,{\rm origin}} &=& {i \over \sqrt{48\pi^3}}
{1\over3^{1/4}} tr[\gamma_1 \lambda_a^5]\nn\\
M_{a(5)}^{2,{\rm origin}} &=& {i \over \sqrt{48\pi^3}}
{3^{1/4}} tr[\gamma_2 \lambda_a^5]\nn\\
M_{a(5)}^{3,{\rm origin}} &=& {i \sqrt{2{\cal V}_3}\over
\sqrt{24\pi^3}} {16^{1/4}} tr[\gamma_3 \lambda_a^5]
\label{MI5sZ6}\eea

The coefficients of $C_I$s are proportional to the coefficients of
$M_I$s (without the traces):
\bea C_{(99,55)}^{k=1}=-4M_{(9,5)}^{1}~,~~~~
C_{(99,55)}^{k=2}=-4M_{(9,5)}^{2}~,~~~~
C_{(99,55)}^{k=3}=-M_{(9,5)}^{3}~. \label{C95sofZ6}\eea
for the various sectors in (\ref{MI9sZ6},\ref{MI5sZ6}).

Now, we can evaluate the symmetric tensor $t_{ijl}$ for the $Z_6$
orientifold using:
\bea t^{Z_N}_{ijl}=\sum_{k=1}^{N-1}\sum_{f}\h_k\left(
M_{i}^{k,f}C_{jl}^{k,f}+
M_{j}^{k,f}C_{li}^{k,f}+M_{l}^{k,f}C_{ij}^{k,f}\right) \ ,
\label{tijl}\eea
where for axionic exchange between D9-D9 and D5-D5
$\eta_1=\eta_2=-\eta_4=-\eta_5=-1$ however, between D9-D5
$\eta_1=\eta_2=-\eta_4=-\eta_5=1$. In all cases $\eta_3=0$.
$M_I$s and $C_I$s are given in (\ref{MI9sZ6}, \ref{MI5sZ6},
\ref{C95sofZ6}).

Using unnormalized $\l$'s ((\ref{lamdas}) without the coefficient
$1/2\sqrt{n_i}$ that normalizes $Tr[\l_i \l_j]=\d_{ij}/2$) we find
perfect agreement with the anomaly matrixes $t_{ijk}$ of the
previous section (\ref{tijZ6}).
We stress that this equation  holds irrespective of the scheme
used in calculating triangle graphs in the effective field theory.

We now evaluate the antisymmetric combination
\bea E^{Z_N}_{ijl}=\sum_{k=1}^{N-1}\sum_{f}\h_k\left(
M_{i}^{k,f}C_{jl}^{k,f}-M_{j}^{k,f}C_{li}^{k,f}\right)\label{Eijl}\eea
that provides the coefficients of the GCS terms, and we find it is
non-zero.
We focus on elements $E_{ijj}=-E_{jij}$ since all other vanish,
$E_{iij}=E_{ijl}=0$:
\be E_{ij}=E_{ijj}=-E_{jij}=\left(
\begin{array}{rrrrrr}
  0 &   36 &-72 &  36 &   0 &-24 \\
-36 &    0 & 72 &   0 & -36 & 24 \\
 24 &  -24 &  0 &  24 & -24 &  0 \\
 36 &    0 &-24 &   0 &  36 &-72 \\
  0 &  -36 & 24 & -36 &   0 & 72 \\
 24 &  -24 &  0 &  24 & -24 &  0
\end{array}\right) \ . \ee
Therefore, in the natural EFT regularization scheme which treats
democratically the anomalous currents, we need GCS terms to cancel
the anomalies in the  $Z_6$ orientifold.

\subsection{$Z_6'$ orbifold}

The orbifold rotation vector is $(v_1,v_2,v_3)=(1,-3,2)/6$. There
is an order two twist ($k=3$) and we must have one set of
D5-branes. Tadpole cancellation then implies the existence of 32
D9-branes and 32 D5-branes, as in the previous example, that we
put together at the origin of the internal space. The Chan--Paton
vectors are
\be V_9=V_5={1\over 12}(1,1,1,1,5,5,5,5,3,3,3,3,3,3,3,3) \ , \ee
implying
\be tr[\gamma_k]=0\ \ \ {\rm for}\ \ k=1,3,5\sp tr[\gamma_2]=-8\sp
tr[\gamma_4]=8\, . \ee
The gauge group has a factor of $U(4)\times U(4)\times U(8)$
coming from the D9-branes and an isomorphic factor coming from the
D5-branes. The massless spectrum is provided in table
\ref{Massless}. The ${\cal N}=1$ sectors correspond to $k=1,5$,
while $k=2,3,4$ are ${\cal N}=2$ sectors.

\subsection*{Anomaly traces}

Normalizing the generators as for the $Z_6$ case, we have for the
mixed anomalies between abelian and non-abelian factors:
\be t_{ia}\equiv Tr[Q_i(T^AT^A)_a]=\left(
\begin{array}{rrrrrr}
 2 & 2 & 8 & -2 & 0 &-4 \\
-2 &-2 & -8 & 0 & 2 & 4 \\
 0 & 0 & 0 &  2 &-2 & 0 \\
-2 & 0 &-4 &  2 & 2 & 8 \\
 0 & 2 & 4 & -2 &-2 &-8 \\
 2 &-2 & 0 &  0 & 0 & 0
\end{array}
\right) \ , \ee
where the columns label again the U(1)s and the rows the
non-abelian factors. We also evaluate the mixed anomalies of
abelian factors $t_{ijk}=Tr[Q_iQ_jQ_k]$ (here we provide again the
$t_{ij}$ (\ref{tij})).
\be t_{ij}=\left(
\begin{array}{rrrrrr}
 48 &   16 & 64 & -16 &   0 &-32 \\
-16 &  -48 &-64 &   0 &  16 & 32 \\
  0 &    0 &  0 &  32 & -32 &  0 \\
-16 &    0 &-32 &  48 &  16 & 64 \\
  0 &   16 & 32 & -16 & -48 &-64 \\
 32 &  -32 &  0 &   0 &   0 &  0
\end{array}\right) \ . \label{tijZ6prime}\ee
In total, there are six U(1)s. Four of them are anomalous and
two are free of 4d anomalies.

\subsection*{Anomalous U(1) masses}

The contribution to the mass matrix is:
\bea {1\over 2}M^2_{aa,ij}&=&-{\sqrt{3}\over
24\pi^3}\left(tr[\gamma_1\lambda_i]
tr[\gamma_1\lambda_j]+tr[\gamma_5\lambda_i]tr[\gamma_5\lambda_j]\right)\nn\\
&-&{(2{\cal V}_2)^{\e_a}\over 8\pi^3}\left(tr[\gamma_2\lambda_i]
tr[\gamma_2\lambda_j]+tr[\gamma_4\lambda_i]tr[\gamma_4\lambda_j]\right)
-{{\cal V}_3\over
3\pi^3}tr[\gamma_3\lambda_i]tr[\gamma_3\lambda_j]~~~~~~~
\eea
where $a=9,5$ and $\e_{9,5}=\pm 1$ respectively, while
\bea {1\over 2}M^2_{95,ij} =&-&{\sqrt{3}\over
48\pi^3}\left(tr[\gamma_1\lambda_i]
tr[\gamma_1\tilde\lambda_j]+tr[\gamma_5\lambda_i]
tr[\gamma_5\tilde\lambda_j]\right. \nn\\
&+& \left. tr[\gamma_2\lambda_i]tr[\gamma_2\tilde\lambda_j]-
tr[\gamma_4\lambda_i]tr[\gamma_4\tilde\lambda_j]\right) -{{\cal
V}_3\over
12\pi^3}tr[\gamma_3\lambda_i]tr[\gamma_3\tilde\lambda_j]\,
\eea

Thus, the unormalized mass matrix has eigenvalues and
eigenvectors:

\bea
\begin{tabular}{lcl}
$m_1^2=6{\cal V}_2$&,& $-A_1+A_2$,\\
$m_2^2={3/ (2{\cal V}_2)}$&,&$ -\tilde A_1+\tilde A_2$,\\
$m_{3,4}^2={5\sqrt{3}+48{\cal V}_3\pm\sqrt{3(25-128 \sqrt{3}{\cal
V}_3+768{\cal V}_3^2)}\over 12}$&,&
$\pm a_{\pm}(A_1+A_2-\tilde A_1-\tilde A_2)-A_3+\tilde A_3$,\\
$m_{5,6}^2={15\sqrt{3}+80{\cal V}_3\pm\sqrt{5(135-384\sqrt{3}{\cal
V}_3+1280{\cal V}_3^2)} \over 12}$&,& $b_{\pm}(A_1+A_2+\tilde
A_1+\tilde A_2)+A_3+\tilde A_3$
\end{tabular}~~~~~~\label{MassesZ61}
\eea
where
\be a_{\pm}={\mp 3+\sqrt{25-128\sqrt{3}{\cal V}_3+768{\cal V}_3^2}
\over 4\sqrt{2} (4\sqrt{3}{\cal V}_3-1)}~~,~~~~~ b_{\pm}={\pm
9\sqrt{3}-\sqrt{5(135-384\sqrt{3}{\cal V}_3 +1280{\cal V}_3^2)}
\over 4\sqrt{2}(20{\cal V}_3-3\sqrt{3})}\, . \ee

Note that the eigenvalues are always positive. They are also
invariant under the T-duality symmetry of the theory ${\cal
V}_2\to 1/4{\cal V}_2$. Thus, all U(1)s become massive,
including the two anomaly free combinations. The reason is that
these combinations are anomalous in six dimensions. Observe
however that in the limit ${\cal V}_3\to 0$, the two linear
combinations that are free of four-dimensional anomalies become
massless. This is consistent with the fact that the
six-dimensional anomalies responsible for their mass cancel
locally in this limit \cite{akr}.

\subsection*{Generalized Chern-Simons terms}
As for the $Z_6$ case, we identify the fixed points and the
couplings of the axions to the branes.

\begin{figure}[h]
\begin{center}
\epsfig{file=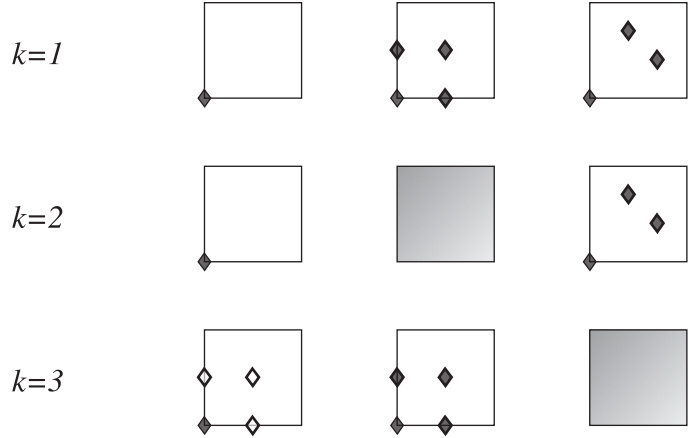,width=90mm}
\end{center}\caption{We denote by
$\blacklozenge$/$\lozenge$ the fixed points on each torus, which
are invariant/related to others by the $Z_6'$ action.
\label{fpZ6prime}}
\end{figure}

The $k=1$ sector provides 12 points fixed under the $Z_6'$ action.
The $k=2$ sector provides 9 points fixed under the $Z_3$ action.
However, the $Z_6'$ action leaves invariant only 3 of them and
relates doublets of the rest. In total there are 3 doublets of
points which are identified under the $Z_6'/Z_3=Z_2$ action.
The $k=3$ sector provides 16 points fixed under the $Z_2$ action.
However, the $Z_6'$ action leaves invariant only 4 of them and
relates triplets of the rest. In total there are 4 triplets of
points identified under the $Z_6'/Z_2=Z_3$ action. In figure
\ref{fpZ6prime} we denote the fixed points under the $Z_6'$
action.

Therefore here the D9 branes couple to axions with:
\bea
M_{a(9)}^{1,\circlearrowleft} &=& {i \over \sqrt{48\pi^3}}
{1\over12^{1/4}} tr[\gamma_1 \lambda_a] \ , \nn\\
M_{a(9)}^{2,\circlearrowleft} &=& {i \sqrt{2{\cal V}_2}\over
\sqrt{24\pi^3}} {1\over9^{1/4}} tr[\gamma_2 \lambda_a]~,~~~~
M_{a(9)}^{2,\rightleftarrows} = {i \sqrt{2{\cal V}_2}\over
\sqrt{24\pi^3}}
{\sqrt{2}\over9^{1/4}} tr[\gamma_2 \lambda_a] \ , \nn\\
M_{a(9)}^{3,\circlearrowleft} &=& {i \sqrt{2{\cal V}_3}\over
\sqrt{24\pi^3}} {1\over16^{1/4}} tr[\gamma_3 \lambda_a]~,~~~~
M_{a(9)}^{3,\rightleftarrows} = {i \sqrt{2{\cal V}_3}
\over\sqrt{24\pi^3}} {\sqrt{3}\over16^{1/4}} tr[\gamma_3
\lambda_a]
 \ , \label{MI9sZ6prime}\eea
where again $\circlearrowleft$ denotes fixed points of the $k$th
sector which are also fixed under the larger $Z_6'$ orbifold
action and $\rightleftarrows$ denotes fixed points which are
related to others by the larger $Z_6'$ orbifold action.

If all D5 branes are at the origin,
\bea
M_{a(5)}^{1,{\rm origin}} &=& {i \over \sqrt{48\pi^3}}
{4^{1/4}\over3^{1/4}} tr[\gamma_1 \lambda_a^5] \ , \nn\\
M_{a(5)}^{2,{\rm origin}} &=& {i \over \sqrt{48\pi^3 {\cal V}_2}}
{3^{1/4}} tr[\gamma_2 \lambda_a^5] \ , \nn\\
M_{a(5)}^{3,{\rm origin}} &=& {i \sqrt{2{\cal V}_3}\over
\sqrt{24\pi^3}} {16^{1/4}} tr[\gamma_3 \lambda_a^5]
\ . \label{MI5sZ6prime}\eea

The coefficients of $C_I$s are proportional to the coefficients of
$M_I$s (without the traces):
\bea C_{(99,55)}^{k=1}=-4M_{(9,5)}^{1}~,~~~~
C_{(99,55)}^{k=2}=-4M_{(9,5)}^{2}~,~~~~
C_{(99,55)}^{k=3}=-M_{(9,5)}^{3}~. \label{C95sofZ6prime}\eea
for the various sectors in (\ref{MI9sZ6prime},\ref{MI5sZ6prime}).
In addition, for axionic exchange between D9-D9 and D5-D5 we have
$\eta_1=-\eta_5=-1$, $\eta_2=\eta_4=0$ however, between D9-D5
$\eta_1=\eta_2=-\eta_4=-\eta_5=1$. In all cases $\eta_3=0$.
Inserting the above to (\ref{tijl}), we evaluate the symmetric
tensor $t_{ijl}$ and find agreement with the anomaly matrix
(\ref{tijZ6prime}).

Similarly, we evaluate the antisymmetric tensor (\ref{Eijl}) for
$Z_6'$:
\be E_{ij}=E_{ijj}=-E_{jij}=\left(
\begin{array}{rrrrrr}
  0 &  -16 &  0 & -16 &   0 & 32 \\
 16 &    0 &  0 &   0 &  16 &-32 \\
 64 &  -64 &  0 & -32 &  32 &  0 \\
-16 &    0 & 32 &   0 & -16 &  0 \\
  0 &   16 &-32 &  16 &   0 &  0 \\
-32 &   32 &  0 &  64 & -64 &  0
\end{array}\right) \ . \ee
Therefore, in the natural regularization scheme which treats
democratically the anomalous currents, we need GCS for the $Z_6'$
as well to cancel the anomalies.

We expect similar couplings to be present in other type of
orientifold models, where anomaly cancellation is taken care by
untwisted axions, like some intersecting / magnetized brane models
\cite{intersecting,magnetized}.


\section{Computation of anomaly diagrams}

 In the following we define: $t^{ijk}=\sum_f[Q^i_f Q^j_f Q^k_f]$.
The triangle amplitude in (\ref{FigureDiagrams}), in momentum space,
is given by
\bea
\G^{ijk}_{\m\n\r}|_{1-loop}=i^3t^{ijk}\int{d^4 p\over (2\pi)^4}
{Tr[\g_\m(\sla{p}+\sla{k_2})\g_\r \sla{p}\g_\n
(\sla{p}-\sla{k_1})\g_5] \over (p+k_2)^2(p-k_1)^2 p^2}
\eea
and can be decomposed according to
\bea
\G^{ijk}_{\m\n\r}|_{1-loop}&=& t^{ijk} [A_1(k_1,k_2) \e_{\m\n\r\s}
k_2^\s + A_2(k_1,k_2) \e_{\m\n\r\s} k_1^\s
+B_1(k_1,k_2) k_{2\n} \e_{\m\r\s\t} k_2^\s k_1^\t\nn\\
&&~~+B_2(k_1,k_2) k_{1\n} \e_{\m\r\s\t} k_2^\s k_1^\t + B_3(k_1,k_2)
k_{2\r} \e_{\m\n\s\t}k_2^\s k_1^\t + B_4(k_1,k_2) k_{1\r}
\e_{\m\n\s\t}k_2^\s k_1^\t ] \nn\\
\label{1loop1}\eea
where $A$'s and $B$'s functions of $k_1,k_2$. In addition to the
triangle diagram in (\ref{FigureDiagrams}), we have to add a similar
triangle diagram with the exchange of $\{k_2,\r \} \Leftrightarrow
\{ k_1,\n\}$. The extra diagram will be similar to the above and the
total result will be twice (\ref{1loop1}) if $A_1(k_1,k_2) = -
A_2(k_2,k_1)$, $B_1(k_1,k_2) = - B_4(k_2,k_1)$, $B_2(k_1,k_2) = -
B_3(k_2,k_1)$.

To evaluate the coefficient functions $A_1,A_2,B_1,B_2,B_3,B_4$, we
use Feynman parametrization
\bea
\G^{ijk}_{\m\n\r}|_{1-loop}&=&i^3t^{ijk}\int d\a d\b
d\g~\d(1-\a-\b-\g) \int{d^4 p \over (2\pi)^4}
{N_{\m\n\r}(p,k_1,k_2)
\over [\a(p+k_2)^2+\b(p-k_1)^2+\g p^2]^3}\nonumber\\
&=&i^3t^{ijk}\int d\a d\b \int{d^4 p \over (2\pi)^4}
{N_{\m\n\r}(p,k_1,k_2)
\over [\a(p+k_2)^2+\b(p-k_1)^2+(1-\a-\b) p^2]^3} \ . 
\eea
We then make the change of variables $\tilde{p}=p+\a k_2- \b k_1$
and redefine back $\tilde{p}\to p$. We thus get
\bea
\G^{ijk}_{\m\n\r}|_{1-loop} &=&i^3t^{ijk}\int d\a d\b \int{d^4p
\over (2\pi)^4}
{N_{\m\n\r}(p,k_1,k_2)
\over [p^2+\a(1-\a)k_2^2+\b(1-\b)k_1^2+2\a\b k_1 k_2]^3}\nonumber\\
&=&i^3t^{ijk}\int d\a d\b \int{d^4p \over (2\pi)^4}
{N_{\m\n\r}(p,k_1,k_2) \over [p^2-P^2+\a k_2^2+\b k_1^2]^3} \ , \eea
where $P=\a k_2-\b k_1$. After the change of variables, the
numerator is :
\bea N_{\m\n\r}& = & Tr[\g_\m(\sla{p}-\sla{P}+\sla{k_2})\g_\r
(\sla{p}-\sla{P})\g_\n
(\sla{p}-\sla{P}-\sla{k_1})\g_5]\nn\\
%
%
&= & -Tr[\g_\m\sla{p}\g_\r
\sla{p}\g_\n(\sla{P}+\sla{k_1})\g_5]
%
-Tr[\g_\m(\sla{P}-\sla{k_2})\g_\r
\sla{p}\g_\n \sla{p}\g_5]
%
- Tr[\g_\m\sla{p}\g_\r \sla{P}\g_\n
\sla{p}\g_5] \nn \\
%
%
&&-
Tr[\g_\m(\sla{P}-\sla{k_2})\g_\r\sla{P}\g_\n(\sla{P}+\sla{k_1})\g_5]
+ \cdots \  \label{traces} \eea
We keep only terms with an even number of $p$'s, which are not
identically zero. Terms with two $p$'s include a logarithmic
divergence, whereas the last term in (\ref{traces}) is convergent.
In dimensional regularization,  using the fact that $p^\m p^\n\to
g^{\m\n} p^2/d$, and $\g^\l \g_\m\g_\l=-(d-2)\g_\m$, $\g^\l \g_\m
\g_\n \g_\r \g_\l=-2\g_\r \g_\n \g_\m +(4-d) \g_\m \g_\n \g_\r$, the
$p^2$ terms can be written as
\bea
&&p^2\Big[{2-d\over d}~(P^\l+k_1^\l+P^\l-k_2^\l)
+{(d-6)\over d}~P^\l\Big]
~Tr[\g_\m\g_\r \g_\n\g_\l\g_5]\nn\\
&=&-{p^2\over d}~\left(-[2-d+(d+2)\a] k_2^\l +[2-d+(d+2)\b]k_1^\l
\right)~(-4i \e_{\m\n\r\l}) \ .
%
\eea
After integration these yield the functions $A_1$ and $A_2$ in
(\ref{1loop1}), which are logarithmically divergent.

The term in the last line in (\ref{traces}) needs no regularization
and can therefore be computed directly in four dimensions $d=4$. The
results is \bea && [(1-\a)P^2-\b k_1^2]~k_2^\l~
Tr[\g_\m\g_\n\g_\r\g_\l\g_5]
-[(1-\b)P^2-\a k_2^2]~k_1^\s~ Tr[\g_\m\g_\n\g_\r\g_\s\g_5]\nn\\
&&+[2\a(\a-1)k_{2\r} -2\a\b k_{1\r}]~k_2^\l k_1^\s~ Tr[\g_\m\g_\n\g_\l\g_\s\g_5]\nn\\
&&+[2\a\b k_{2\n}- 2 \b(\b-1) k_{1\n}]~k_2^\l k_1^\s~
Tr[\g_\m\g_\r\g_\l\g_\s\g_5]\nn
\eea

Now we can perform the integrals on $p$ : \be \int {d^4p \over
(2\pi)^4}{1\over (p^2+M^2)^3} = {1\over 32\pi^2  M^2} \ , \ee
whereas \be \int {d^4p \over (2\pi)^4}{p^2\over (p^2+M^2)^3} \quad
 \ee
is logarithmically divergent and thus scheme dependent.
On the contrary the finite coefficients $B$'s of (\ref{1loop1}) can
be determined unambiguously and read
\bea B_1(k_1,k_2)=-{i \over 8 \pi^2} \int_0^1 d\a d\b {2 \a\b\over
\a k_2^2+\b k_1^2-(\a k_2-\b k_1)^2}
\ , \nn\\
B_2(k_1,k_2)=-{i \over 8 \pi^2} \int_0^1 d\a d\b {2 \b (1-\b)\over
\a k_2^2+\b k_1^2-(\a k_2-\b k_1)^2}
\ , \nn\\
B_3(k_1,k_2)=-{i \over 8 \pi^2}\int_0^1 d\a d\b {-2 \a (1-\a)\over
\a k_2^2+\b k_1^2-(\a k_2-\b k_1)^2}
\ , \nn\\
B_4(k_1,k_2)=-{ i \over 8 \pi^2}\int_0^1 d\a d\b {-2 \a\b\over \a
k_2^2+\b k_1^2-(\a k_2-\b k_1)^2} \ . \label{Bs}\eea
Notice that: $B_1(k_1,k_2)= - B_4(k_2,k_1) = - B_4(k_1,k_2)$,
$B_2(k_1,k_2)=-B_3(k_2,k_1)$. The functions $A_i$, {\it a priori}
logarithmically divergent, will be determined by imposing
renormalization conditions on the three-point function.

\subsection{All diagrams}

Consider the three-point function of (abelian) gauge-bosons $\langle
A_{\mu}^i(k_3) A_{\nu}^j(k_1) A_{\rho}^k(k_2)\rangle$ in momentum
space. There are three contributions to this three-point function,
linear in the momenta. One comes from the the irreducible CS-like
vertex and gives a contribution \be \langle A_{\m}^i A_{\n}^j
A_{\r}^k \rangle_{GCS} \ = \  -
{\epsilon^{\mu\nu\rho}}_{\sigma}\left[ E_{ijk}
~k_2^{\s}+E_{kij}~k_1^{\s}+E_{jki}~k_3^{\s}\right] \ , \ee where we
have used antisymmetry in the first two indices.

The second contribution comes from axion-vector mixing terms and PQ
couplings of the axions to $F-\tilde F$. We have the vertex \be
a^I(k_3)A^i_{\mu}(k_1)A^j_{\mu}(k_2)\to
2i~C^I_{ij}~{\epsilon^{\mu\nu}}_{\rho\s}~k_2^{\rho} k_3^{\s} \ee
where there is a factor of two coming from each field strength and a
factor of 1/2 from the definition of the dual. The mixing term is
\be a^I(k)A^i_{\mu}(k)\to - M_I^i \ k^{\mu} \ee and  the axion
propagator \be a^I(k)a^J(k)\to {iG^{IJ}\over k^2} \ , \ee where
axion indices are raised and lowered with the axion metric, to be
taken to be canonical $G_{IJ} =\delta_{IJ}$ in what follows.
Performing all six contractions we obtain \be \langle A_{\m}^i
A_{\n}^j A_{\r}^k \rangle_{\rm contact} = - \left[ C_I^{ij} M_I^k
{\epsilon^{\mu\nu}}_{\s\s'}~k_3^{\s} k_1^{\s'}~{k_2^{\rho}\over
k_2^2} + C_I^{ik} M_I^j~{\epsilon^{\mu\rho}}_{\s\s'}~k_3^{\s}
k_2^{\s'}~{k_1^{\nu}\over k_1^2} \right.\ee
$$
\left. +C_I^{jk} M_I^i {\epsilon^{\nu\rho}}_{\s\s'}~k_1^{\s}
k_2^{\s'}~{k_3^{\mu}\over k_3^2}\right] \ .
$$

Combining all the contributions one gets
\bea \G^{ijk}_{\m\n\r}=\G^{ijk}_{\m\n\r}|_{1-loop}
+\G^{ijk}_{\m\n\r}|_{axion} +\G^{ijk}_{\m\n\r}|_{CS} \ , \eea
where
\bea
\G^{ijk}_{\m\n\r}|_{1-loop}&=& t^{ijk} [ A_1 \e_{\m\n\r\s} k_2^\s +
A_2 \e_{\m\n\r\s} k_1^\s
+B_1 k_{2\n} \e_{\m\r\s\t} k_2^\s k_1^\t+B_2 k_{1\n} \e_{\m\r\s\t} k_2^\s k_1^\t\nn\\
&&~~+B_3 k_{2\r} \e_{\m\n\s\t}k_2^\s k_1^\t + B_4 k_{1\r}
\e_{\m\n\s\t} k_2^\s k_1^\t] \ , \\
\G^{ijk}_{\m\n\r}|_{axion}&=& -M^i_I C^{jk}_I \left({k_{3\m} \over
k_3^2}\right) \e_{\n\r\s\t} k_2^\s k_1^\t
-M^j_I C^{ki}_I \left({k_{1\n} \over k_1^2}\right) \e_{\r\m\t\s}
k_2^\s k_3^\t
-M^k_I C^{ij}_I \left({k_{2\r} \over k_2^2}\right) \e_{\m\n\t\s}
k_3^\s k_1^\t \nn\\
&=& -M^i_I C^{jk}_I {-(k_{1\m}+k_{2\m}) \over
(k_1+k_2)^2}\e_{\n\r\s\t} k_2^\s k_1^\t
+M^j_I C^{ki}_I {k_{1\n} \over k_1^2} \e_{\m\r\s\t} k_2^\s k_1^\t
-M^k_I C^{ij}_I {k_{2\r} \over k_2^2} \e_{\m\n\s\t} k_2^\s k_1^\t
\nn\\
\\
\G^{ijk}_{\m\n\r}|_{CS}&=& -E^{ij,k} \e_{\m\n\r\s} k_2^\s
-E^{jk,i} \e_{\n\r\m\s} k_3^\s -E^{ki,j} \e_{\r\m\n\s} k_1^\s\nn\\
&=& -(E^{ij,k}-E^{jk,i}) \e_{\m\n\r\s} k_2^\s
-(E^{ki,j}-E^{jk,i}) \e_{\m\n\r\s} k_1^\s \ .
\eea
Imposing total Bose symmetry of the amplitude\footnote{ In order to
check the symmetry of the amplitude at the interchange of the
external gauge bosons, we can use the identity:
\bea (k_{1\m} + k_{2\m}) \e_{\n\r\s\t} k_2^\s k_1^\t &=& -(k_{2\n} +
k_{1\n}) \e_{\r\m\s\t}k_2^\s k_1^\t
-(k_{1\r} + k_{2\r}) \e_{\m\n\s\t} k_2^\s k_1^\t\nn\\
&&+ \epsilon_{\mu\nu\rho\sigma}[k_2^2 k_1^\sigma - k_1^2 k_2^\sigma
- k_1. k_2 (k_2^\sigma - k_1^\sigma)] \ .
\eea}, the appropriate anomaly conditions for the triangle diagrams
turn out to be
\bea \left. \begin{array}{l}
k_1^\n\G^{ijk}_{\m\n\r}|_{1-loop}=-t^{ijk} {C_A \over 3} \e_{\m\r\s\t} k_2^{\s} k_1^{\t}\\
k_2^\r\G^{ijk}_{\m\n\r}|_{1-loop}=t^{ijk} {C_A \over 3} \e_{\m\n\s\t} k_2^{\s} k_1^{\t}\\
-(k_1^\m+k_2^\m)\G^{ijk}_{\m\n\r}|_{1-loop} =t^{ijk} {C_A \over 3}
\e_{\n\r\s\t} k_2^{\s} k_1^{\t} \end{array} \right)
~~~\longrightarrow~~~
\left. \begin{array}{l}
A_1 + k_1\cdot k_2 B_1+k_1^2B_2=-{C_A \over 3} \nn \\
A_2+k_2^2 B_3+k_1\cdot k_2 B_4= {C_A \over 3} \nn \\
A_1-A_2 ={C_A \over 3}\end{array} \right. \eea
where $C_A$ is the standard coefficient of the axial anomaly. In
this scheme, we can express the ambiguous amplitudes $A_1,~A_2$ (a
priori logarithmically divergent) in terms of the finite
$B_1,~B_2,~B_3,~B_4$ given in (\ref{Bs})\footnote{We can change the
scheme by redefining the scheme-dependent coeff. $A_i$. For example,
if we require $k_2^\r\G^{ijk}_{\m\n\r}|_{1-loop}=0,~
k_1^\n\G^{ijk}_{\m\n\r}|_{1-loop}=0,~
(k_1^\m+k_2^\m)\G^{ijk}_{\m\n\r}|_{1-loop} = t^{ijk} C_A$, by
choosing the anomaly to be contained in the third current, this can
be done by the redefinitions  $(A_1,A_2)\to
(A_1-C_A/3,A_2+C_A/3).$}.

Requiring the vanishing of the total anomaly we then find
\bea \left. \begin{array}{l}
k_1^\n\G^{ijk}_{\m\n\r}=0\\
k_2^\r\G^{ijk}_{\m\n\r}=0\\
(k_1^\m+k_2^\m)\G^{ijk}_{\m\n\r} =0 \end{array} \right)
~~~\longrightarrow~~~
\left. \begin{array}{r}
t^{ijk}(A_1+k_1\cdot k_2 B_1+k_1^2B_2)+M_I^j C_I^{ki}-E^{ij.k}+E^{jk.i}=0 \nn \\
t^{ijk}(A_2+k_2^2B_3+k_1 \cdot k_2 B_4)-M_I^k C_I^{ij}-E^{ki.j}+E^{jk.i}=0 \nn \\
t^{ijk}(A_1-A_2)-M_I^i C_I^{jk}+E^{ki.j}-E^{ij.k}=0 \end{array}
\right. \nn \\
\label{vanishing} \eea
We can also express the anomaly cancellation conditions in the form
\bea t^{ijk}C_A&=&M_I^k C_I^{ij}+M_I^j C_I^{ki}+M_I^i C_I^{jk} \ , \nn \\
E^{ij.k}&=&{1\over 3}(M_I^i C_I^{jk}-M_I^j C_I^{ki}) \ , \nn \\
E^{jk.i}&=&{1\over 3}(M_I^j C_I^{ki}-M_I^k C_I^{ij}) \ , \nn \\
E^{ki.j}&=&{1\over 3}(M_I^k C_I^{ij}-M_I^i C_I^{jk}) \ .
\label{E=MC}\eea
From the above equations it is easy to notice that for a gauge
invariant model, the totaly symmetric part of $M_I^{(i} C_I^{jk)}$
cancels the anomaly (which is a totaly symmetric tensor $C_A
t^{ijk}$) and the antisymmetric part $M_I^{[i} C_I^{j]k}$ cancels
$E^{[ij]k}$. Therefore, one can construct a gauge invariant model
with anomalous fermion content and axions (without the GCS terms),
one can construct a gauge invariant model without chiral fermion
content ($t^{ijk}=0$) \cite{gcs4} but we cannot construct a model
with anomalous fermions and GCS terms without axions.


\end{document}